\documentclass[preprint,12pt]{elsarticle}

\makeatletter
\def\ps@pprintTitle{%
	\let\@oddhead\@empty
	\let\@evenhead\@empty
	\def\@oddfoot{\centerline{\thepage}}%
	\let\@evenfoot\@oddfoot}
\makeatother

\usepackage{lineno}
\usepackage{hyperref}
\usepackage{amsfonts}
\usepackage{amssymb}
\usepackage{url} %this package should fix any errors with URLs in refs.
\usepackage{graphicx}
\usepackage{amsmath}
\usepackage{bm}
\usepackage{color}
\usepackage{csquotes}
\usepackage{float}
\usepackage{gensymb}
\usepackage{mathtools}
\usepackage{multirow}
\modulolinenumbers[5]

\journal{Journal of Hydrology}

\biboptions{authoryear}

\begin{document}
%\linenumbers
\begin{frontmatter}
%\begin{linenumbers}

\title{On the relation between parameters and discharge data for a lumped karst aquifer model}


\author[hrbm]{Daniel Bittner\corref{correspondingauthor}}
\cortext[correspondingauthor]{Corresponding author} 
\ead{daniel.bittner@tum.de}

\author[nm]{Mario Teixeira Parente}
\author[nm]{Steven Mattis}
\author[nm]{Barbara Wohlmuth}
\author[hrbm,uibk]{Gabriele Chiogna}

\address[hrbm]{Chair of Hydrology and River Basin Management, Technical University of Munich, Arcisstr. 21, 80333 Munich, Germany}
\address[nm]{Chair for Numerical Mathematics, Technical University of Munich, Boltzmannstraße 3, 85748 Garching, Germany}
\address[uibk]{Institute of Geography, University of Innsbruck, Innrain 52, 6020 Innsbruck, Austria}

\begin{abstract}

Hydrological models of karst aquifers are often semi-distributed, and physical processes such as infiltration and spring discharge generation are described in a lumped way. 
Several works have previously addressed the problems associated with the calibration of such models, highlighting in particular the issue of model parameter estimation and model equifinality. 
In this work, we investigate the problem of model calibration using the active subspace method, a novel tool for model parameter dimension reduction. 
We apply the method to a newly proposed hydrological model for karst aquifers, LuKARS (Land use change modeling in KARSt systems), to investigate if the active subspace framework identifies catchment-specific characteristics or if the results only depend on the chosen model structure. 
Therefore, we consider four different case studies, three synthetic and one real case (Kerschbaum springshed in Waidhofen a.d. Ybbs, Austria), with varying hydrotope distributions and properties. 
We find that both the hydrotope area coverage and the catchment characteristics (here: water storage and discharge properties of the hydrotopes) have major impacts on parameter sensitivities.
While model parameters are similarly informed in scenarios with less varying catchment characteristics, we find significant differences in parameter sensitivities when the applied hydrotopes were different from each other. 
Our results show that the active subspace method can be used to investigate the relation between the model structure, the area of a hydrotope, the physical properties of a catchment and the discharge data.
Finally, we successfully effectively reduce the parameter dimensions of the LuKARS model for the Kerschbaum case study using the active subspace method.
The model with reduced parameter dimensions is able to reproduce the observed impacts of land use changes in the Kerschbaum springshed, highlighting the robustness of the hydrotope-based modeling approach of LuKARS and its applicability for land use change impact studies in karstic systems.

\end{abstract}

\begin{keyword}
Rainfall-discharge modeling, Karst hydrology, Sensitivity analysis, Acitve subspaces, Land use change
\end{keyword}

\end{frontmatter}

\section{Introduction}

\label{S:Intro}
%Parameter calibration is a fundamental step to be performed before applying any hydrological model \citep{beven1992future}. 
%A large amount of literature describes several different methods to estimate model parameters \citep{tarantola2005inverse,aster2011parameter,kitanidis2014principal}. 

%A recently developed method, called the active subspace method \citep{Constantine.2014}, quantifies a weight for each model parameter, thus, providing a measure of the amount of information associated with a given parameter and a given calibration dataset. 
%To the best of our knowledge, a previous application to a lumped rainfall-discharge model for karstic systems does not exist, but it is valuable since most of the lumped parameters are not measurable and, hence, it is difficult to quantify how the measurements used for model calibration inform the model parameters.\par
Hydrological models are commonly applied to investigate hydrological processes \citep{fleury2007modelling} and the hydrological impacts of lan use changes in defined catchments \citep{sarrazin2018v2karst}. 
Distributed hydrologic models are the preferred tools for these purposes as they can provide reasonable physical representations of hydrological properties and processes \citep{Henson.2018,Chen.2018}. 
Those models require detailed information about surface and subsurface properties of an area of interest. 
However, in the particular case of karst aquifers, that information often cannot be obtained due to the highly complex and heterogeneous subsurface structure of karstic systems \citep{ladouche2014semi,jukic2009groundwater}.
This considerable lack of spatially distributed information about subsurface flow in karst systems makes conceptual rainfall-discharge models suitable tools to predict karst spring discharge, since they allow to drastically simplify the description of the functioning of the subsurface system \citep{Hosseini.2017}. 
%Model predictions about spring discharge are therefore affected by significant uncertainties \citep{Chen.2018,Henson.2018}. 
However, finding an acceptable model representation is not trivial since it is difficult to constrain the model parameter range according to field observations and measurements. 
The general trade-off is two-fold. 
Firstly, a severe simplification of a complex karst system may lead to an underrepresentation of dominant processes, such as groundwater recharge, conduit flow and land use dependent infiltration on karst systems.
Disregarding these processes also means that the model cannot be used for some hydrological investigations, such as the effect of land use change on spring water discharge. 
However, a low-dimensional parameter space, i.e. 4 to 6 parameters \citep{Jakeman.1993}, can reduce parameter uncertainties and model equifinality \citep{Hartmann.2017}.
Secondly, considering more complexity in the modeling framework can improve the conceptual representation of the regarded karst system \citep{Hartmann.2018}.
However, the parameters of a more complex lumped model often cannot be adequately calibrated since they may not be sufficiently informed by the measured discharge data \citep{Hartmann.2014}. 
We call a model parameter informed if the objective function measuring deviation from observed data is sensitive to it. 
 \par
%Most applied karst rainfall-discharge models use a minimum of 6 parameters \citep{Hartmann.2017}. 
Therefore, the sensitivity and the identification of the parameters which are informed by the observations used for model calibration is a primary concern for the community dealing with karst modeling. 
Different approaches were recently proposed to reduce model structure and parameter uncertainties for rainfall-discharge models applied in karst hydrology.
\citet{Hartmann.2013} developed an evaluation strategy to identify the most valid model structure while finding a balance between model structure, parameter identification and model performance for a karst system of interest. 
Moreover, \citet{Hartmann.2017} presented an approach to consider hydrochemical data to constrain parameter ranges of the VarKarst model \citep{Hartmann.2013b}. 
Further multi-objective calibration frameworks were proposed by \citet{Moussu.2011}, \citet{Mazzilli.2013} and \citet{mudarra2019combining}, using the autocorrelation of the discharge, ground-based gravity measurements and dye tracer concentrations to better constrain model outputs.   
However, the high dimensionality of the parameter space remains a challenge for karst hydrologic models, since it is difficult to quantify how and which model parameters are informed by the measurements used to calibrate the model. 

\par  
A pertinent example of a model with a high-dimensional parameter space is the LuKARS (Land use change modeling in KARSt systems) model that was recently proposed by \citet{Bittner.2018} to simulate the hydrological impacts of land use changes in a karst aquifer. 
The LuKARS model is based on the implementation of hydrotopes, which are distinct landscape units characterized by homogeneous hydrological properties as a result of similar land use and soil types \citep{Koeck.2012,arnold1998large}.
Each hydrotope in LuKARS has 7 calibration parameters.
For one scenario, \citet{Bittner.2018} set up the LuKARS model for the Kerschbaum spring recharge area in Waidhofen a.d. Ybbs (Austria) including four hydrotopes.
This high-dimensional parameter space makes the model prone to structural and parameter uncertainties.
A suitable approach is therefore needed to perform a global sensitivity analysis of the full parameter space.
Moreover, it is worth investigating if it is possible to reduce the dimension of the parameter space, i.e. to identify a limited number of parameters that are sufficient for calibrating the model.
A recently developed method, called the active subspace method \citep{Constantine.2014}, when applied to inverse problems identifies directions in parameter space that are most informed by data. 
These parameter directions contain valuable information about parameter sensitivities and how model parameters are related to each other for a given objective function. 
So far, the active subspace method was used in different research studies, mostly applied to model reduction contexts \citep{Constantine.2015,parente2018efficient,Jefferson.2015}.
\citet{TeixeiraParente.2019} provide a detailed overview of the mathematical details of the active subspace method as applied to a lumped karst aquifer model.

\par
In this work, in contrast, we present how the active subspace method can be used to gain a better understanding of model functioning and which hydrological implications may be derived from these results.
In particular, we want to investigate if the results of the active subspace method only depend on the model structure or also on the specific discharge behavior of the hydrotopes in a catchment. 
We perform active subspace analysis on the recently proposed hydrotope-based LuKARS model \citep{Bittner.2018} for three synthetic scenarios and the Kerschbaum spring case study considering in each case a 21-dimensional parameter space. 
Our particular research objectives are (i) to investigate different synthetic hydrotope scenarios to derive how hydrotope modifications and varying areas covered by each hydrotope affect the results of the dimension reduction stategry, (ii) to investigate which and how the hydrotope parameters of the LuKARS model for the Kerschbaum spring are informed by discharge data, and (iii) to discuss the hydrological implications of the relationship between hydrotope parameters and karst spring discharge data under varying catchment conditions.

\section{Methods}
\subsection{LuKARS model}
\label{sec:lukars}
The general idea of the LuKARS modeling approach is to perform land use change impact studies in a karstic environment by implementing the model in areas with homogeneous infiltration conditions as distinct hydrological response units called hydrotopes \citep{Bittner.2018}. 
%The core of the LuKARS model is a non-linear hysteretic transfer function \citep{Tritz.2011} which is used to implement each hydrotope. 
A hydrotope $i$ represents a bucket that has three discharge components: the quickflow component (conduit flow, $Q_{\text{hyd},i}$ [L$^3$T$^{-1}$]), a secondary spring discharge ($Q_{\text{sec},i}$ [L$^3$T$^{-1}$]) and the recharge ($Q_{\text{is},i}$ [L$^3$T$^{-1}$]). 
$Q_{\text{hyd},i}$ is considered a hydrotope-specific quickflow component occurring in preferential flow paths (e.g. subsurface conduits). 
The quickflow bypasses the baseflow storage $B$ and is directly transferred to the spring outlet.
The quickflow starts once a hydrotope-specific storage threshold ($E_{\text{max},i}$) has been reached and is responsible for the fast reaction of a spring discharge to rainfall and snowmelt events. 
$Q_{\text{sec},i}$ integrates all flow components that do not arrive at the investigated karst spring and that are transferred outside the regarded recharge area, i.e. secondary spring discharge and overland flow \citep{Tritz.2011}. 
$Q_{\text{is},i}$ represents the discharge from hydrotope $i$ to the underlying baseflow storage $B$, representing the process of groundwater recharge. 
%The baseflow storage then controls the baseflow contribution ($Q_b$) to the spring by a linear transfer function. 
 \par

LuKARS solves the following discrete balance equations for each time step \(n\):

%\begin{equation}
%\label{eq_hyd_balance}
%\frac{dE_i}{dt}=\begin{cases}
%S_i-\frac{Q_{sec,i}+Q_{is,i}+Q_{hyd,i}}{a_i} 
%& \text{if $E_i>0$}\\
%0 & \text{if $E_i=0$}
%\end{cases}
%\end{equation}
\begin{equation}
\label{eq_hyd_discr}
E_{i,n+1}=\max[0,E_{i,n}+(S_{i,n}-\frac{Q_{\text{hyd},i,n}+Q_{\text{sec},i,n}+Q_{\text{is},i,n}}{a_i}) \, \Delta t] 
\end{equation}
is the balance equation solved for each hydrotope, where $E_{i}$ indicates the water level [L] in hydrotope $i$.
\(S_i\) is the hydrotope-specific sink and source term as a mass balance of precipitation, snow melt, evapotranspiration and interception. 
Interception is calculated using estimates provided by \citet{dvwk1996ermittlung}. 
A simple temperature index model \citep{martinec1960degree} is used to model snow melt and snow retention in the model. 
Then, evapotranspiration is considered using the method proposed by \citet{thornthwaite1948approach}.
The absolute area covered by a respective hydrotope is given by \(a_i\) [L$^2$].

%\begin{equation}
%\label{eq_baseflow_balance}
%\frac{dE_b}{dt}=\begin{cases}
%\frac{\Sigma(Q_{is,i})+Q_b}{A}
%& \text{if ${E_b}>0$}\\
%0 & \text{if ${E_b}=0$}
%\end{cases}
%\end{equation}

\begin{equation}
\label{eq_baseflow_discr}
E_{\text{b},n+1}=\max[0,E_{\text{b},n}+(\frac{\Sigma(Q_{\text{is},i,n})-Q_{\text{b},n}}{A}) \, \Delta t] 
\end{equation}
is the balance equation for the baseflow storage $B$, where $E_\text{b}$ indicates the water level [L] in the baseflow storage and \(\Sigma(Q_{\text{is},i})\) [L$^3$T$^{-1}$] are the cumulative flows from all hydrotopes to the baseflow storage. 
$Q_\text{b}$ [L$^3$T$^{-1}$] represents water that is transferred from the storage \(B\) to the spring and simulates a baseflow contribution from the phreatic aquifer system to the spring discharge. 
The variable \(A\) [L$^2$] stands for the entire recharge area. 
%The discretized forms of \ref{eq_hyd_balance} and \ref{eq_baseflow_balance}, as given in \ref{eq_hyd_discr} and \ref{eq_baseflow_discr}, are solved for each time step \(n\):

The discharge terms are computed as follows:
\begin{equation}
\label{eq_hyd_discharge}
Q_{\text{hyd},i,n}=a_i \, \frac{k_{\text{hyd},i}}{l_{\text{hyd},i}} \, \varepsilon_{n}[\frac{\max(0,E_{i,n}-E_{\text{min},i})}{E_{\text{max},i}-E_{\text{min},i}}]^{\alpha_{i}}
\end{equation}
\begin{equation}
\label{eq_sec_discharge}
Q_{\text{sec},i,n}=a_i \, k_{\text{sec},i} \, \max(0,E_{i,n}-E_{\text{sec},i}) 
\end{equation}
\begin{equation}
\label{eq_is_discharge}
Q_{\text{is},i,n}=a_i \, k_{\text{is},i} \, E_{i,n}
\end{equation}
\begin{equation}
\label{eq_baseflow_discharge}
Q_{\text{b},n}=A \, k_\text{b} \, E_{\text{b},n}
\end{equation}
\(E_{\text{max},i}\) [L] and \(E_{\text{min},i}\) [L] represent the upper and lower storage thresholds of the hydrotope \(i\). \(E_{\text{sec},i}\) [L] is the hydrotope-specific activation level for a secondary spring discharge. 
\(k_{\text{sec},i}\) [LT$^{-1}$], \(k_{\text{is},i}\) [LT$^{-1}$] and \(k_\text{b}\) [LT$^{-1}$] are the specific discharge parameters for \(Q_{\text{sec},i}\) [L$^3$T$^{-1}$], \(Q_{\text{is},i}\) [L$^3$T$^{-1}$] and \(Q_\text{b}\) [L$^3$T$^{-1}$], respectively. 
\(k_{\text{hyd},i}\) [L$^2$T$^{-1}$] represents the specific discharge parameter for the quickflow of a hydrotope and \(l_{\text{hyd},i}\) [L] is the mean distance of hydrotope \(i\) to the adjacent spring, thus accounting for the relative location of the same hydrotope types in a specific recharge area. 
The ratio between \(k_{\text{hyd},i}\) and \(l_{\text{hyd},i}\) represents the hydrotope discharge coefficient and $\alpha_{i}$ is a hydrotope-specific exponent of the quickflow. 
The dimensionless connectivity/activation indicator $\varepsilon$ is defined as follows:

\begin{equation}
\label{eps_1}
\varepsilon_{n+1}=\text{0 if}\begin{cases}
\text{$\varepsilon_n=0$ \& $E_{i,n+1}<E_{\text{max},i}$ or}\\
\text{$\varepsilon_n=1$ \& $E_{i,n+1} \leq E_{\text{min},i}$}
\end{cases}
\end{equation}

\begin{equation}
\label{eps_2}
\varepsilon_{n+1}=\text{1 if}\begin{cases}
\text{$\varepsilon_n=0$ \& $E_{i,n+1} \geq E_{\text{max},i}$ or}\\
\text{$\varepsilon_n=1$ \& $E_{i,n+1}>E_{\text{min},i}$}
\end{cases}
\end{equation}

The hydrotope implementation is based on the classification of similar soil and land use areas within a catchment and most parameters in LuKARS cannot be directly obtained by field measurements. 
It is important to note, that the conceptual character of LuKARS and the large number of calibration parameters make this approach prone to uncertainties.
%Hence, there is a general need to investigate the informativeness of each single parameter, as well as to quantify parameter uncertainty. 
%A deeper understanding of the way how model parameters are informed by data can significantly improve the reliability of water resources predictions with LuKARS. \par
A more detailed description for how these processes were considered in the framework of LuKARS is provided in \citet{Bittner.2018}.

\subsection{Physical relationship between hydrotopes and their parameters}
\label{sec:physical_param_rel}

A typical setup of LuKARS for a defined recharge area includes different hydrotopes. 
Each hydrotope is said to show characteristic hydrological responses to precipitation events determined by its soil and land use properties. 
To what extent one particular hydrotope response contributes to the total catchment response depends on the area covered by a considered hydrotope in a defined recharge area.
On the one hand, hydrotopes having coarse-grained and shallow soils should have a high contribution to the quickflow and groundwater recharge (having a high connectivity to fractures and conduits). 
Moreover, the possibility that these hydrotopes become  may dry after a long period without any precipitation should be allowed. 
On the other hand, the parametrization of hydrotopes with more fine-textured and deep soils should allow to show slow and minor contributions to the quickflow and the groundwater recharge but be able to store a certain water volume.
This means, that a given parameter set of a hydrotope with a small storage volume (e.g. a shallow, coarse-grained soil) needs to be interpreted in relation to the parameters applied to a second hydrotope with a higher storage volume (e.g. a thicker, more fine-grained soil). 
These physical constraints have to be met to accept a set of hydrotope parameters in LuKARS. 
%It is important to note that Hyd Q is not part of model calibration since the respective areas are drained on the surface and do not contribute to the total spring discharge.

\subsection{Identification of the parameters informed during inverse modeling}
\label{S:active_subsp}
To identify the most dominant parameters which are informed by the data used for model calibration, we exploit the active subspace method, a recently developed method for dimension reduction \citep{constantine2014computing,constantine2015active,Constantine.2014}.
The method identifies important directions in a parameter space, i.\,e., directions that, on average, dominantly change a given function of interest $f$.
Using these directions, it can also provide global sensitivity metrics for the parameters which is explained below.
Recently, active subspaces have been used to reduce the dimension of parameter spaces in high-dimensional Bayesian inverse problems \citep{constantine2016accelerating,cortesi2017forward,parente2018efficient,TeixeiraParente.2019}.

In the following, we summarize the basic theory of active subspaces for finding dominant directions in a parameter inference problem.
We consider the situation in which $\bm{d}$ represents the observed time series data that we aim at modeling (e.g. spring discharge in our case).
The map $\mathcal{G}$ represents our forward LuKARS model with a given set of input parameters.
The definition of the calibration parameters is provided in Section~\ref{relations}.
We assume $\bm{\eta} \sim \mathcal{N}(\bm{0}, \Gamma)$ is zero-centered Gaussian noise with covariance matrix~$\Gamma$, such that:

\begin{equation}
\bm{d} = \mathcal{G}(\bm{x})  + \bm{\eta}.
\end{equation}

The function of interest $f$, for which we want to identify dominant directions, is a corresponding data misfit function, denoted by $f_{\bm{d}}$, from the context of Bayesian inverse problems \cite{stuart2010inverse}, i.\,e.,
\begin{equation}
\label{eq:misfit}
f_{\bm{d}}(\bm{x}) \coloneqq \frac{1}{2}\| \bm{d} - \mathcal{G}(\bm{x}) \|_{\Gamma}^2 \coloneqq \frac{1}{2}\| \Gamma^{-1/2} \left(\bm{d} - \mathcal{G}(\bm{x}) \right) \|_{2}^2.
\end{equation}

To identify important directions of $f=f_{\bm{d}}$, one looks at eigenpairs of the symmetric positive semi-definite $n \times n$ matrix 
\begin{align}
\label{eq:C}
\bm{C} &\coloneqq %\expct{\rho}{\grad f \grad f \tr} \\
\int{\nabla f(\bm{x}) \nabla f(\bm{x})^{\top} \rho(\bm{x}) \, d\bm{x}} = \bm{W} \Lambda \bm{W}^{\top},
\end{align}
where $\rho$ is a given probability density function, and $f$ is assumed to be continuous and differentiable on the support of $\rho$, additionally having square-integrable derivatives with respect to $\rho$.
The matrix $\bm{W}=[\bm{w}_1,\dots,\bm{w}_n]$ contains the eigenvectors $\bm{w}_i$ of $\bm{C}$, and the matrix $\Lambda=\text{diag}(\lambda_1,\dots,\lambda_n)$ is a diagonal matrix with the eigenvalues on the diagonal.
The eigenvalues of $\bm{C}$ provide information about the average sensitivity of $f$ in the direction of the corresponding eigenvector, since it holds that
\begin{equation}
\lambda_i = \bm{w}_i^\top \bm{C} \bm{w}_i = \int{(\bm{w}_i^\top \nabla f(\bm{x}))^2 \rho(\bm{x}) \,d\bm{x}}.
\end{equation}
If the eigenvalues decay quickly, there are directions in the space of unknown parameters where $f$ varies much more, on average, than other directions, i.e. these directions are significantly more informed by the data.
The span of the eigenvectors associated with significant eigenvalues is called the active subspace.
If an active subspace is identified, it can be used to effectively reduce the dimension of the parameter space for inversion, potentially greatly reducing the computational cost.
Simultaneously, not all directions are well-informed by the data, so new data, to which parameters contributing to these directions are sensitive, would be needed to improve the estimation of the remaining parameters.

In practice, $\bm{C}$ can be approximated by a finite Monte Carlo sum, i.\,e.
\begin{equation}
\label{eq:mc_sum}
\bm{C} \approx \frac{1}{N}\sum_{j=1}^{N}{\nabla f(\bm{x}_j)\nabla f(\bm{x}_j)}^{\top} = \bm{\tilde{W}} \tilde{\Lambda} \bm{\tilde{W}}^{\top}
\end{equation}
for $N\in\mathbf{N}$ and samples $\bm{x}_j$ distributed according to $\rho$.
The matrices $\bm{\tilde{W}}=[\bm{\tilde{w}}_1,\dots,\bm{\tilde{w}}_n]$ and $\tilde{\Lambda}=\text{diag}(\tilde{\lambda}_1,\dots,\tilde{\lambda}_n)$ denote perturbations to their exact counterpart from Eq.~\ref{eq:C}.
A relatively small number of gradient samples (in our case about 250) is required to calculate a sufficiently accurate approximation of the eigenvectors and eigenvalues of $\bm{C}$ (see \citet{Constantine.2015}).
Gradients can be calculated by using methods such as adjoint approaches, finite differences, and radial basis functions (see e.\,g., \citet{plessix2006review}, \citet{mai2003approximation}).

The approximated eigenpairs $(\bm{\tilde{w}}_i, \tilde{\lambda}_i)$ can be used to compute global sensitivity metrics \citep{constantine2017global} comparable with more familiar metrics like Sobol indices \citep{sobol2001global}.
The components of the vector~$\bm{s}$ consisting of the sensitivity values are defined as
\begin{equation}
s_i = \sum_{j=1}^{n} \lambda_j(\bm{w}_j)_i^2, \quad i = 1,\dots,n.
\end{equation}
More compactly, we can write
\begin{equation}
\bm{s} = (\bm{W}\circ\bm{W}) \bm{\lambda},
\end{equation}
where $\circ$ denotes elementwise multiplication and $\bm{\lambda}=(\lambda_1,\dots,\lambda_n)$ is the vector of eigenvalues.

\section{Application}
\label{S:Field}
\begin{figure}[htb]
	\centering
	\includegraphics[width=\textwidth]{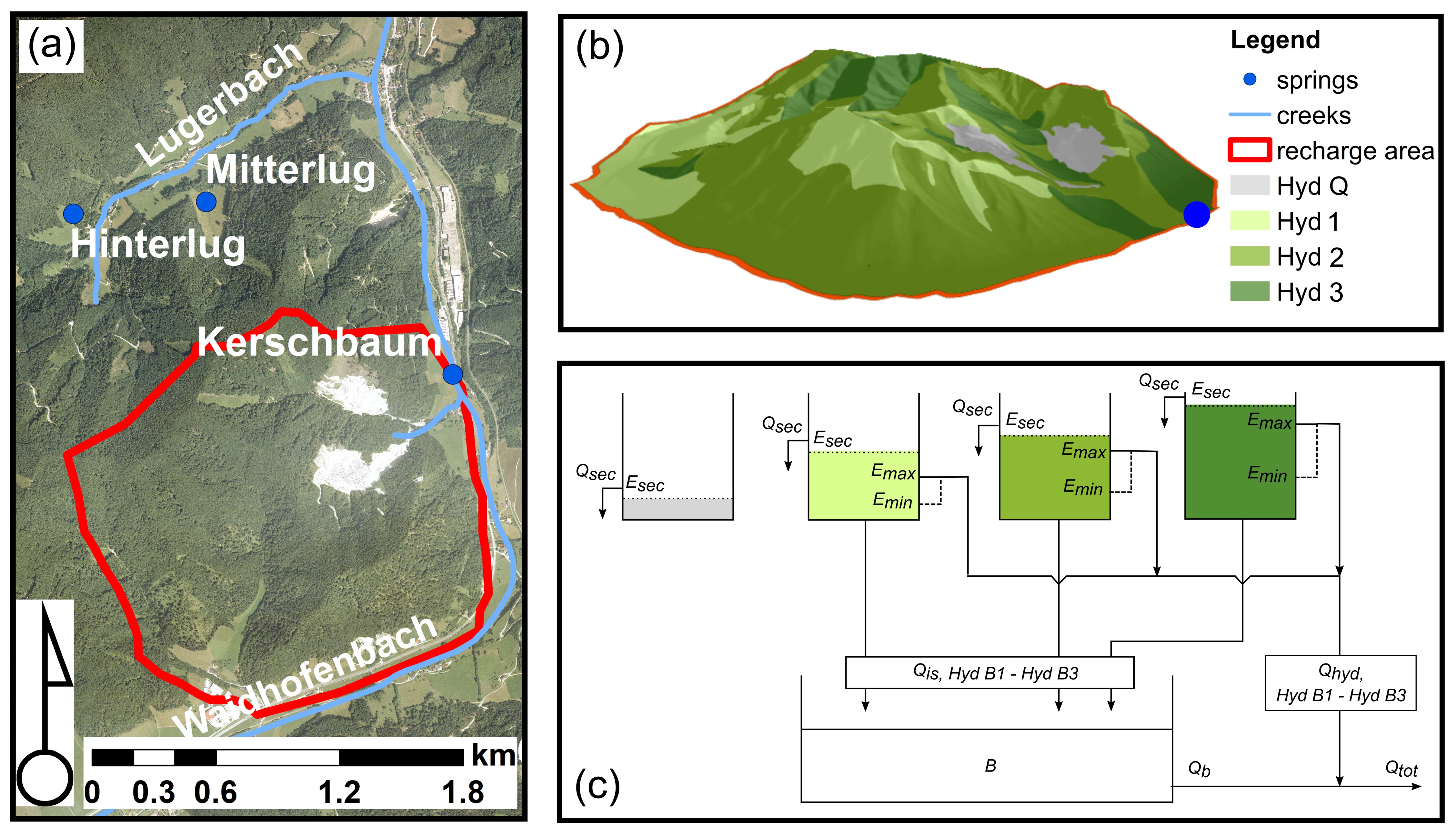}
	\caption{Overview of the study area Waidhofen a.d. Ybbs (Austria) and the conceptualization of the Kerschbaum spring recharge area. (a) The location and the boundary of the Kerschbaum spring recharge area in the study area. The orthophoto shows that the predominant landcover is forest. (b) Pseudo-3-dimensional view on the topography of the Kerschbaum recharge area including the hydrotope classification according to \citet{Koeck.2012}. (c) Conceptual model of the recharge area as implemented in LuKARS, considering each hydrotope as a distinct storage unit. The different sizes of the tanks show the different storage capacities of the respective hydrotopes (not to scale).}
	\label{fig:study_area}
\end{figure}

Our case study is the Kerschbaum karst springshed located 10km south of Waidhofen a.d. Ybbs (Austria) (Fig. \ref{fig:study_area}a). 
On a regional scale, this pre-alpine catchment is part of the eastern foothills of the Northern Calcareous Alps with altitudes ranging between 415m and 969m a.s.l. 
The area is further characterized by a warm-moderate regional climate with a mean annual temperature of 8$^{\circ}$C and a mean annual precipitation of 1379mm. The annual distribution of the precipitation is bimodal with maxima during the summer (June and July) and winter periods (December and January), being indicative for the relevance of rainfall and snowfall-related processes in the study area \citep{Bittner.2018}. \par
Several spring outlets exist in the study area of which some are used for the local water supply of the community of Waidhofen a.d. Ybbs. 
Here, the Kerschbaum spring represents the most important one with a mean annual discharge of 34ls\textsuperscript{-1}. 
Prior hydrogeological investigations revealed that the Kerschbaum spring is fed by the karst aquifer of the dolomitic bedrock \citep{Hacker.2003}, which is part of the Triassic Main Dolomite strata of the Northern Calcareous Alps. 
The small-scale recharge area comprises a surface area of 2.5km\textsuperscript{2} and is predominantly covered by beech forests. 
A more detailed description of the study area can be found in \citet{Bittner.2018} and \citet{SheikhyNarany.2018}. 

\subsection{Hydrotope scenarios}
We develop three synthetic hydrotope scenarios for which we investigate the interdependencies between the calibration parameters of a hydrotope and the observed spring discharge. 
These synthetic scenarios are based on the properties of the hydrotopes present in the Kerschbaum springshed (Fig. \ref{fig:study_area}b). 
For this reason, we first provide a description of the Kerschbaum hydrotopes (real case) before we introduce the synthetic scenarios in this section. 
The conceptual implementation of the hydrotopes into LuKARS is illustrated for the real case in Fig. \ref{fig:study_area}c).
The input time series (precipitation and temperature) along with the measured and synthetic discharge time series are shown in Fig. \ref{fig:ts_scenarios}a) and b), respectively.

\subsubsection{Real case}
\label{S:real_case}
Following the hydrotope classification performed by \citet{Koeck.2012}, four major hydrotopes are present in the Kerschbaum spring recharge area. 
During the investigated period of this research study (2006-2008), about 4\% of the recharge area were used for dolomite mining activities, represented by hydrotope Hyd Q in Fig. \ref{fig:study_area}b).
This space increased to about 7\% after the year 2009. 
Three more hydrotopes were classified in the recharge area, which are Hydrotope 1 (denoted by Hyd~1), Hydrotope 2 (denoted by Hyd~2) and Hydrotope 3 (denoted by Hyd~3) (shown in Fig. \ref{fig:study_area}b). 
Originally, the hydrotopes are named after their dominant land cover according to \citet{Koeck.2012}. 
Hyd~1 (\textit{Bluegrass-Beech Forest hydrotope}) is characterized by shallow and coarse-grained soils, being indicative for a low storage capacity and covers 13\% of the Kerschbaum recharge area. 
Hyd~2 (\textit{White Sedge-Beech Forest hydrotope}) covers 56\% of the considered reacharge area and has more fine-grained soils with moderate thicknesses. 
Finally, Hyd~3 (\textit{Christmas Rose-Beech Forest hydrotope}) represents the second largest hydrotope with 27\% of the coverage and is dominated by mostly loam textured soils with more elevated thicknesses and has the highest storage capacity as compared to Hyd~1 and Hyd~2. 

According to the general parameter relationships of LuKARS described in \ref{sec:physical_param_rel}, the following physical parameter constraints count for the Kerschbaum spring model:
\begin{equation}
\label{eq:ineq_constr}
\begin{gathered}
k_{\text{hyd},{1}} \geq k_{\text{hyd},{2}} \geq k_{\text{hyd},{3}}, \\
E_{\text{min},1} \leq  E_{\text{min},2} \leq  E_{\text{min},3}, \\
E_{\text{max},1} \leq  E_{\text{max},2} \leq  E_{\text{max},3}, \\
\alpha_{1} \geq \alpha_{2} \geq \alpha_{3}, \\
k_{\text{is},1} \geq  k_{\text{is},2} \geq  k_{\text{is},3}, \\
k_{\text{sec},1} \geq  k_{\text{sec},2} \geq  k_{\text{sec},3}, \\
E_{\text{sec},1} \leq  E_{\text{sec},2} \leq  E_{\text{sec},3}.
\end{gathered}
\end{equation}
Based on our experiences made during the intensive manual calibration procedure performed in the framework by \citet{Bittner.2018}, we were able to define reasonable parameter ranges of each model parameter and each hydrotope. 
Those parameter ranges are used as prior intervals for the active subspace framework in the presented research and are shown in Table \ref{param_intervals}.  

\begin{table}[htb]
	\caption{Overview of the model parameters, their physical units and descriptions as well as the lower bound (lb) and upper bound (ub) of the prior intervals used for the active subspace method.}
	\centering
	\begin{tabular}{lllll}
		\hline\noalign{\smallskip}
		Parameter & unit & description & lb & ub \\
		\hline\hline\noalign{\smallskip}
		$k_{\text{hyd},1}$	& [L$^{2}$T$^{-1}$]	& discharge parameter for $Q_{\text{hyd},1}$ 	& 9           & 900         \\
		$E_{\text{min},1}$	& [L]	& min storage capacity Hyd~1    		& 10 		& 50          \\
		$E_{\text{max},1}$	& [L]	& max storage capacity Hyd~1		  	& 15       	& 75          \\
		$\alpha_{1}$& [-]	& quickflow exponent of Hyd~1  			& 0.7     	& 1.6         \\
		$k_{\text{is},1}$	& [LT$^{-1}$] 	& discharge parameter for $Q_{\text{is},1}$   		& 0.002       & 0.2         \\
		$k_{\text{sec},1}$	& [LT$^{-1}$] 	& discharge parameter for $Q_{\text{sec},1}$  		& 0.0095      & 0.95        \\
		$E_{\text{sec},1}$	& [L] 	& activation level for $Q_{\text{sec},1}$  		& 25 		& 70          \\
		\hline\noalign{\smallskip}
		$k_{\text{hyd},2}$ & [L$^{2}$T$^{-1}$]	& discharge parameter for $Q_{\text{hyd},2}$  	& 8.5         & 850         \\
		$E_{\text{min},2}$ 	& [L] 	& min storage capacity Hyd~2  		& 40        & 80          \\
		$E_{\text{max},2}$	& [L]	& max storage capacity Hyd~2  		& 80        & 160          \\
		$\alpha_{2}$& [-]	& quickflow exponent of Hyd~2  			& 0.5       & 1.3         \\
		$k_{\text{is},2}$	& [LT$^{-1}$] 	& discharge parameter for $Q_{\text{is},2}$    		& 0.00055     & 0.055       \\
		$k_{\text{sec},2}$	& [LT$^{-1}$] 	& discharge parameter for $Q_{\text{sec},2}$   		& 0.0023      & 0.23        \\
		$E_{\text{sec},2}$	& [L] 	& activation level for $Q_{\text{sec},2}$  		& 130       & 220         \\
		\hline\noalign{\smallskip}
		$k_{\text{hyd},3}$	& [L$^{2}$T$^{-1}$]	& discharge parameter for $Q_{\text{hyd},3}$  	& 7.7         & 770         \\
		$E_{\text{min},3}$	& [L] 	& min storage capacity Hyd~3  		& 75        & 120         \\
		$E_{\text{max},3}$	& [L]	& max storage capacity Hyd~3  		& 155       & 255         \\
		$\alpha_{3}$& [-]	& quickflow exponent of Hyd~3  			& 0.2       & 0.7         \\
		$k_{\text{is},3}$	& [LT$^{-1}$] 	& discharge parameter for $Q_{\text{is},3}$    		& 0.00025     & 0.025       \\
		$k_{\text{sec},3}$	& [LT$^{-1}$]	& discharge parameter for $Q_{\text{sec},3}$    	& 0.0015      & 0.15        \\
		$E_{\text{sec},3}$	& [L]	& activation level for $Q_{\text{sec},3}$    	& 320       & 450         \\
		\hline
	\end{tabular}
	\label{param_intervals}
\end{table}

\subsubsection{Synthetic scenarios}
\begin{figure}[htb]    
	\centering
	\includegraphics[scale=0.52]{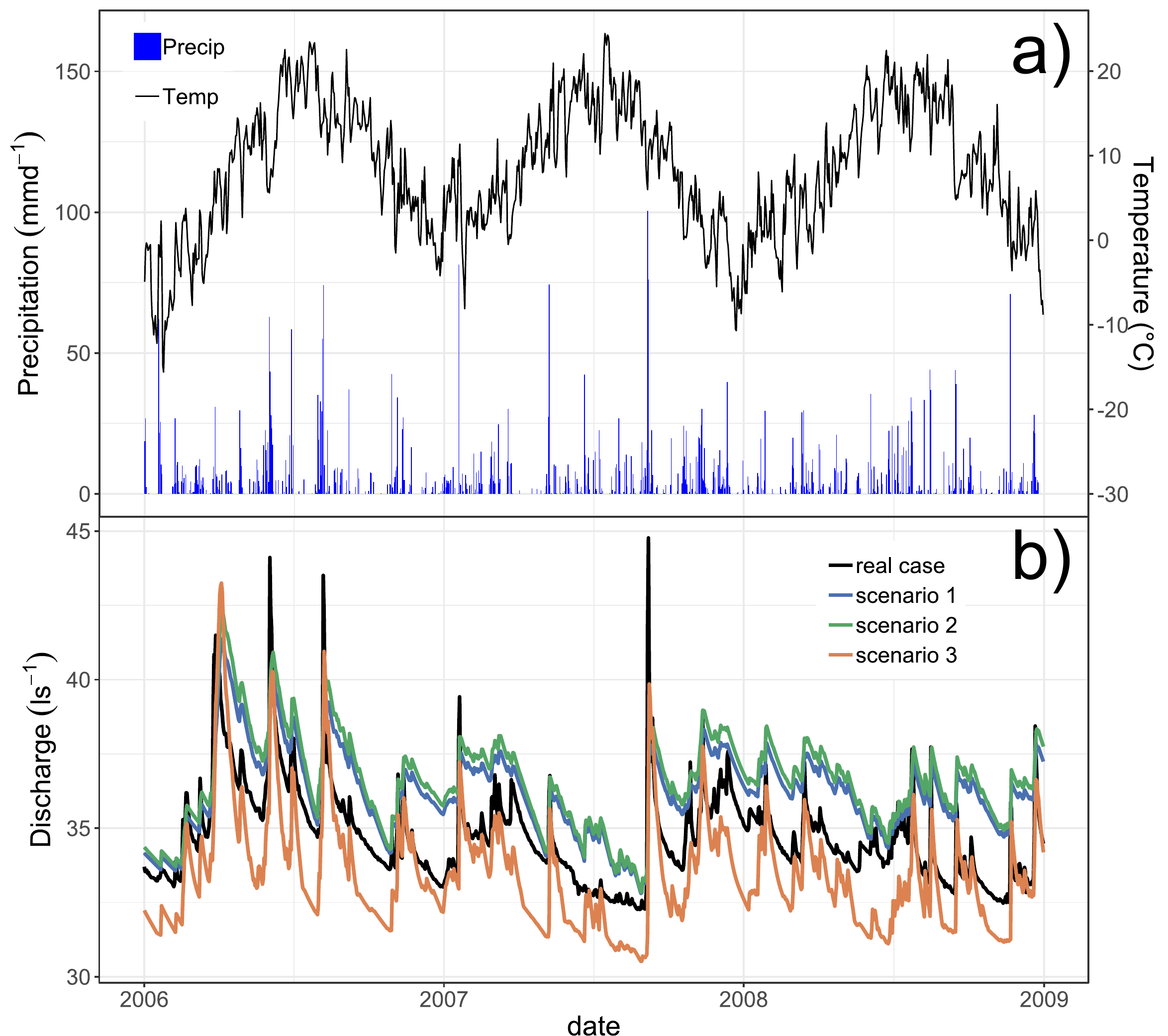}
	\caption{a) The precipitation and temperature time series measured at the Mitterlug spring and b) the spring discharge time series of the Kerschbaum spring for the real case (measured) and the 3 synthetic scenarios.}
	\label{fig:ts_scenarios}
\end{figure}

The synthetic model scenarios are generated based on the hydrotope properties of Hyd~1 to Hyd~3 as implemented for the Kerschbaum spring recharge area (see Section \ref{S:real_case}). 
This means, that we increase or decrease the space of one hydrotope at the expense of another one and/or that we completely replace the properties of one hydrotope with those of another one present in the Kerschbaum recharge area. 
The space of the total recharge area remains constant in all scenarios. 
Moreover, we did not change the space covered by Hyd Q (quarries) and kept its 4\% coverage during the model calibration period from 2006 to 2008. 
For the evaluation of these scenarios, we generate synthetic spring discharge time series of each scenario using the calibrated hydrotope parameters provided by \citet{Bittner.2018} (Fig. \ref{fig:ts_scenarios}). 
Together with the real case scenario, we have a total of four hydrotope models to be evaluated. All scenarios are summarized in Table \ref{table_scenarios} and described in detail in the following: \par

Scenario 1 (All Hyd~2 with same areas) consists of 3 hydrotopes plus Hyd~Q. 
In this synthetic hydrotope scenario, all three hydrotopes are assumed to have the same properties and parameter ranges as Hyd~2 in the real case model (see Table \ref{param_intervals}). 
This means, we consider all hydrotopes to represent distinct landscape units with silty to sandy soils of moderate thicknesses. 
Moreover, all hydrotopes are even in terms of spatial share, meaning that each hydrotope covers 32\% of the recharge area. 
From a hydrological point of view, this scenario leads to a reduced discharge variability as compared to the real case.  
Scenario 1 could be implemented considering only one hydrotope covering 96\% of the recharge area. 
However, our particular aim of this scenario is to investigate the interactions between model parameters and observation data in case of homogeneous hydrotope properties with even areas.
We want to see if the active subspace method recognizes that this scenario has the characteristics of only one hydrotope. 
Therefore, we expect that no preference will be given to a specific hydrotope for what concerns the sensitivity of the hydrotope parameters. We hypothesize that the same parameters will be similarly informed in all hydrotopes.  

\par 
Scenario 2 (All Hyd~2) is similar to scenario 1, since all 3 considered hydrotopes have the same properties and parameter ranges as Hyd~2 in the real case model. 
However, those 3 hydrotopes have the same spatial share as Hyd~1, Hyd~2 and Hyd~3 in the real case scenario, which are 13\%, 56\% and 27\%, respectively. 
Similar to scenario 1, this change implies a reduction in discharge variability as can be seen in Fig. \ref{fig:ts_scenarios}.
However, the reason why the time series of scenario 1 and 3 do not match is related to the model parameter $l_{\text{hyd}}$ (mean distance of a hydrotope to the spring), which was not modified in the framework of this study.
Our particular purpose here is to investigate how different spatial coverages affect the sensitivities of the hydrotope parameters, while having the same physical properties in all hydrotopes.
Therefore, we hypothesize that the parameters of the hydrotope with the largest area coverage (Hyd~2) will be the most sensitive ones and most informed by the discharge data. 
Further, we want to compare the results of scenario 2 with those obtained from scenario 1 to prove the effect of the area coverages of the hydrotopes on the parameter sensitivities.   \par 

In scenario 3 (Real case with different spaces), we maintain the physical properties and parameter ranges of all hydrotopes as they exist in the real case LuKARS model for the Kerschbaum spring recharge area. 
The specific feature of scenario 2 is that we changed the areas covered by each hydrotope, leading to the following modifications: 
Hyd~1 covers 30\% (instead of 13\%), Hyd~2 covers 26\% (instead of 56\%) and Hyd~3 covers 40\% (instead of 27\%) of space in the Kerschbaum spring recharge area. These changes lead to a more pronounced discharge variability since the space of the most dynamic hydrotope, Hyd~1, increased significantly. 
Here, we want to analyse the interdependencies between parameter sensitivities and the specific hydrological behavior (i.e. discharge variability) a regarded hydrotope displays. 
We expect that the parameters of Hyd~1, although it is not the largest hydrotope, will be most informed by the discharge data because it shows the highest discharge variability (discharge dynamics). 
Given the high discharge dynamics displayed by Hyd~1, we assume that its parameters are more sensitive as compared to those of Hyd~2 and Hyd~3. 
These results will further be compared to the real case Kerschbaum model. \par 

\begin{table}[htb]
	\caption{Overview of the characteristics of the synthetic hydrotope scenarios (1-3) and the real case (rc).}
	\centering
	\begin{tabular}{p{0.5cm}p{2.5cm}p{6.5cm}p{3.5cm}}
		\hline\noalign{\smallskip}
		No. & name & description & area coverage \\
		\hline\hline\noalign{\bigskip}
		1 & All Hyd~2 with same areas	& Same like scenario 1 but with same areas covered by each hydrotope  	& Hyd~1 (32\%), Hyd~2 (32\%), Hyd~3 (32\%)        \\
		\noalign{\bigskip}	
		2 & All Hyd~2	& Physical properties of Hyd~1, Hyd~2 and Hyd~3 are equal to Hyd~2 of the real case 	& Hyd~1 (13\%), Hyd~2 (56\%), Hyd~3 (27\%)      \\	
		\noalign{\bigskip}
		3 & Real case with different spaces	& Physical properties of all hydrotopes are like in the real case scenario but with different areas   		& Hyd~1 (30\%), Hyd~2 (26\%), Hyd~3 (40\%)  \\
		\noalign{\bigskip}
		rc & Real case	& Original hydrotope properties of Hyd~1, Hyd~2 and Hyd~3  	& Hyd~1 (13\%), Hyd~2 (56\%), Hyd~3 (27\%)        \\
		\hline
	\end{tabular}
	\label{table_scenarios}
\end{table}

\subsection{Setup of calibration parameters}
\label{relations}
Since the model parameters for each hydrotope have to meet the inequality constraints shown in Eq.~\eqref{eq:ineq_constr}, they are not independent from a statistical point of view.
However, the active subspace framework prefers independent parameters.
Thus, we introduce new synthetic parameters which are called calibration parameters from now on.
The model parameters, also called physical parameters, for hydrotope $i$ are $k_{\text{hyd},i}$ (discharge parameter for $Q_{\text{hyd},i}$), $E_{\text{min},i}$ (minimum storage capacity Hyd $i$), $E_{\text{max},i}$ (maximum storage capacity Hyd $i$), $\alpha_{i}$ (quickflow exponent of Hyd $i$), $k_{\text{is},i}$ (discharge parameter for $Q_{\text{is},i}$), $k_{\text{sec},i}$ (discharge parameter for $Q_{\text{sec},i}$), $E_{\text{sec},i}$ (activation level for $Q_{\text{sec},i}$), where the index $i=1,2,3$ indicates a specific hydrotope.
It is worth distinguishing between the 7 aforementioned model parameters for each hydrotope and the related calibration parameters described in the following.

We need to introduce three types of non-normalized calibration parameters.
\begin{enumerate}
\item The $k_*$ values are calibrated on a log scale. Therefore, we define
\begin{equation}
	k^\text{log}_* = \log(k_*)
\end{equation}
for each $k_* \in \{k_{\text{hyd},{i}}, k_{\text{is},i}, k_{\text{sec},i} \}$, $i=1,2,3$.

\item For $i=1,2,3$, since $E_{\text{min},i} \leq E_{\text{max},i}$, parameter samples for $E_{\text{max},i}$ would be dependent on samples $E_{\text{min},i}$. Hence, we write $ E_{\text{max},i} = E_{\text{min},i}+\Delta E_i$ for new (non-normalized) calibration parameters $\Delta E_i$.
This means that the parameter $E_{\text{max},i}$ is "replaced" by $\Delta E_i$ to get back independence.

\item Similar to point 2), the physical constraints from Eq.~\eqref{eq:ineq_constr} lead to the introduction of new (non-normalized) calibration parameters that mimic the difference between values of two successive hydrotopes.
\end{enumerate}
We make sure that these constraints are met and that values are chosen such that the corresponding model parameters lie in the respective specified intervals from Table~\ref{param_intervals}.
In particular, we write for $i=2,3$
\begin{align}
\begin{split}
k^\text{log}_{\text{hyd},{i}} &= k^\text{log}_{\text{hyd},{i,\text{lb}}} + \Delta k^\text{log}_{\text{hyd},{(i-1,i)}}(\min\lbrace k^\text{log}_{\text{hyd},{i,\text{ub}}},k^\text{log}_{\text{hyd},{i-1}} \rbrace - k^\text{log}_{\text{hyd},{i,\text{lb}}}), \\
E_{\text{min},i} &= \max\left\lbrace  E_{\text{min},i-1}, E_{\text{min},i,\text{lb}} \right\rbrace \\ &\qquad + \Delta E_{\text{min},(i-1,i)} \left( E_{\text{min},i,\text{ub}} - \max\left\lbrace  E_{\text{min},i-1}, E_{\text{min},i,\text{lb}} \right\rbrace\right), \\
\alpha_{i} &= \alpha_{i,\text{lb}} + \Delta\alpha_{(i-1,i)} \left(\min\left\lbrace \alpha_{i,\text{ub}},\alpha_{i-1} \right\rbrace - \alpha_{i,\text{lb}}\right), \\
k^\text{log}_{\text{is},i} &= k^\text{log}_{\text{is},i,\text{lb}} + \Delta k^\text{log}_{\text{is},(i-1,i)} (\min\lbrace  k^\text{log}_{\text{is},i,\text{ub}}, k^\text{log}_{\text{is},i-1} \rbrace -  k^\text{log}_{\text{is},i,\text{lb}}), \\
k^\text{log}_{\text{sec},i} &=  k^\text{log}_{\text{sec},i,\text{lb}} + \Delta k^\text{log}_{\text{sec},(i-1,i)} (\min\lbrace  k^\text{log}_{\text{sec},i,\text{ub}}, k^\text{log}_{\text{sec},i-1} \rbrace -  k^\text{log}_{\text{sec},i,\text{lb}}), \\
E_{\text{sec},i} &= \max\left\lbrace  E_{\text{sec},i-1}, E_{\text{sec},i,\text{lb}} \right\rbrace \\ &\qquad + \Delta E_{\text{sec},(i-1,i)} \left( E_{\text{sec},i,\text{ub}} - \max\left\lbrace  E_{\text{sec},i-1}, E_{\text{sec},i,\text{lb}} \right\rbrace\right),
\end{split}
\end{align}
where lower and upper bounds of each interval for the model parameters are denoted by subscripts ${}_{\text{lb}}$ and ${}_{\text{ub}}$ specified in Table~\ref{param_intervals}.
The parameters with a triangle ($\Delta$) are new calibration parameters, taking values in [0,1] and replacing corresponding model parameters on the left hand side of the equal sign.
Note that new synthetic parameters were not necessary for the first hydrotope ($i=1$) since there are no differences in value to a preceding hydrotope.

Eventually, all non-normalized calibration parameters are normalized, i.\,e., they are mapped to the interval [-1,1].
The normalized parameters are denoted with a bar above their name and form the final 21-dimensional vector $\bm{x}$ of calibration parameters, i.\,e.,
\begin{align}
\label{eq:calib_param}
\begin{split}
\bm{x} = (& \bar{k}^\text{log}_{\text{hyd},1} , \bar{E}_{\text{min},1},\Delta \bar{E}_{1},\alpha_{1}, \bar{k}_{\text{is},1}, \bar{k}_{\text{sec},1}, \bar{E}_{\text{sec},1}, \\
& \Delta \bar{k}^\text{log}_{\text{hyd},{(1,2)}},\Delta \bar{E}_{\text{min},(1,2)},\Delta \bar{E}_2,\Delta\bar{\alpha}_{(1,2)}, \\ &\hspace{1.5cm} \Delta \bar{k}_{\text{is},(1,2)},\Delta \bar{k}_{\text{sec},(1,2)},\Delta \bar{E}_{\text{sec},(1,2)}, \\
& \Delta \bar{k}^\text{log}_{\text{hyd},{(2,3)}},\Delta \bar{E}_{\text{min},(2,3)},\Delta \bar{E}_3,\Delta\bar{\alpha}_{(2,3)}, \\ &\hspace{1.5cm} \Delta \bar{k}_{\text{is},(2,3)},\Delta \bar{k}_{\text{sec},(2,3)},\Delta \bar{E}_{\text{sec},(2,3)})^{\top} \in \mathbf{R}^{21}.
\end{split}
\end{align}

We want to emphasize that the following results, containing parameter sensitivities, are computed w.r.t. the normalized calibration parameters.
However, the interpretations for the original model parameters are still valid.
For example, claiming that $\alpha_3$ is sensitive has a different meaning than saying that $\alpha_1$ is, because $\alpha_3$ can only be chosen dependently on $\alpha_1$ and $\alpha_2$ to satisfy the inequality constraints from Eq.~\eqref{eq:ineq_constr}.
Hence, sensitivity of $\alpha_3$ actually means that the difference to $\alpha_2$ (and $\alpha_1$ indirectly) is sensitive which is exactly what we model with the introduction of synthetic difference parameters denoted with a triangle ($\triangle$).
It is worth mentioning that samples for the dependent model parameters (gained by translating from the corresponding samples for calibration parameters), as e.\,g., $\alpha_3$, are no longer uniformly distributed, since a sum of uniformly distributed random variables no longer follows a uniform distribution.
This is a consequence of the physically motivated inequality constraints and, hence, acceptable.

The focus of the presented research study lies on investigating the hydrotope parameters.
Therefore, the discharge coefficient of the baseflow storage, $k_\text{b}$, is used as calibrated in the work of \citet{Bittner.2018} and is not included in this parameter study. 
Moreover, the parameter constraints shown in section~\ref{S:real_case} and the described setup only hold for the real case scenario and scenario 3. 
Since scenarios 1 and 2 are implemented based on the properties of Hyd~2 from the real case, having the same physical characteristics with different area coverages, there is no need to introduce parameter relationships between different hydrotopes.

\section{Results}
\label{S:Results}
The water works Waidhofen a.d. Ybbs indicated a noise level of 5\% for the measured discharge time series of the Kerschbaum spring. Hence, we considered the same noise level for the synthetic data we generated for scenario 1 to 3.

\subsection{Synthetic scenarios}

\begin{figure}[htbp]
	\centering
	\includegraphics[width=\textwidth]{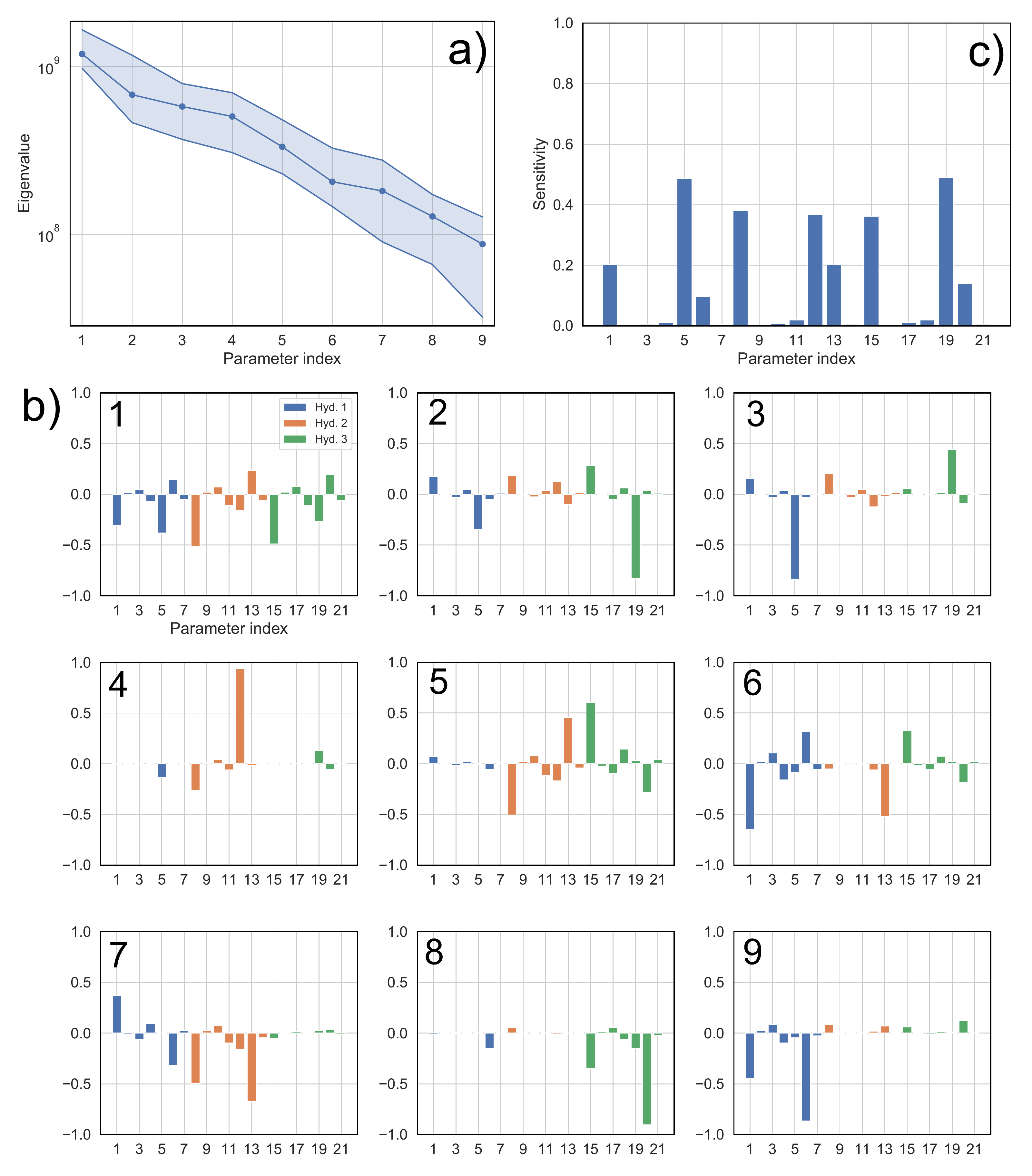}
	\caption{Results of the active subspace method applied to scenario 1 of the Kerschbaum spring LuKARS model. a) The eigenvalue decay with respect to the first 9 eigenvectors. b) The parameter sensitivities of all 21 calibration parameters. c) The first 9 eigenvectors of the 21-dimensional parameter space.}
	\label{fig:scenario_1}
\end{figure}

We show the results of scenario 1 in Fig. \ref{fig:scenario_1}. 
Fig. \ref{fig:scenario_1}a) displays the decay of the first 9 (out of 21) eigenvalues on a logarithmic scale.
The eigenvalue decay helps to identify spectral gaps that point towards the existence of a $n$-dimensional subspace. 
Although we observe small gaps after the first and fourth eigenvalue, these gaps are not pronounced enough, and thus do not point towards the existence of 1-dimensional or 4-dimensional active subspaces.
A spectral gap is considered significant if the decay from one eigenvalue to another is about one order of magnitude.  
Since no significant gap was identified, we consider all eigenvectors lying within the range of one order magnitude of the eigenvalue decay (here: 9). 
The corresponding eigenvectors, including the model parameters that contribute to each eigenvector, are shown in Fig. \ref{fig:scenario_1}b). 
The 21 components of each vector represent the 21 calibration parameters of the model. 
The model parameters associated with each component of the calibration parameter vector $\bm{\vec{x}}$ are indicated in Table \ref{param_intervals}. 
Each hydrotope is characterized by 7 independent parameters. 
Therefore, the components 1-7 are inherent to Hyd~1, the components 8-14 represent Hyd~2 and the components 15-21 Hyd~3. 
Loosely speaking, the model parameters that appear in the eigenvectors corresponding to the highest eigenvalues are most informed by the measured discharge data and suggest high parameter sensitivities in the respective directions of the parameter space.  
It is interesting to see that all hydrotope parameters of Hyd~1 - Hyd~3 appear in the first eigenvector (Fig. \ref{fig:scenario_1}b). 
Moreover, the more pronounced parameter contributions in eigenvector 1 are attributed to the same parameters in each hydrotope, which are the $k_\text{hyd}$, $k_\text{is}$ and $k_\text{sec}$ parameters (parameters 1, 5, 6, 8, 11, 12, 15, 19 and 20). 
The same pattern can be observed when looking at all shown eigenvectors, where the $k_*$ parameters of the three hydrotopes are dominant.
Moreover, all hydrotopes are similarly informed by the discharge data.
The parameters contributing most to the 9 eigenvectors also appear as the most sensitive parameters in the global sensitivity metrics (Fig. \ref{fig:scenario_1}c). 
Comparable to the parameters having pronounced contributions in the first eigenvector, the most sensitive parameters in scenario 1 are the $k_\text{hyd}$, $k_\text{is}$ and $k_\text{sec}$ parameters of Hyd~1 to Hyd~3. \par 

\begin{figure}[htb]
	\centering
	\includegraphics[width=\textwidth]{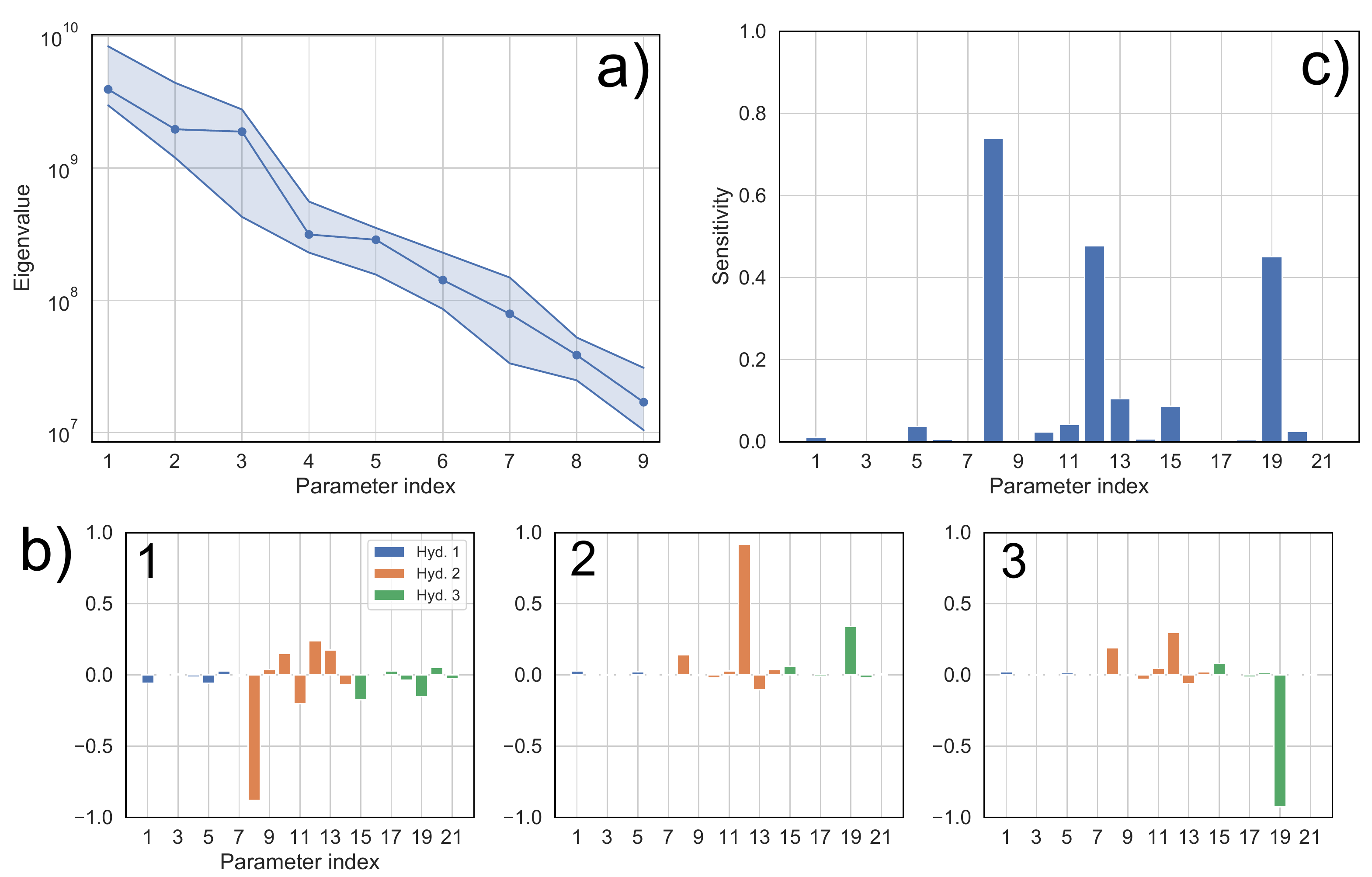}
	\caption{Results of the active subspace method applied to scenario 2 of the Kerschbaum spring LuKARS model. a) The eigenvalue decay as related to the first 9 eigenvectors. b) The parameter sensitivities of all 21 calibration parameters. c) The first 3 eigenvectors of the 21-dimensional parameter space.}
	\label{fig:scenario_2}
\end{figure}

Next, the results of scenario 2 are shown in Fig. \ref{fig:scenario_2}. 
Here, we can observe a pronounced spectral gap after the third eigenvalue, indicating the possible existence of a 3-dimensional active subspace (Fig. \ref{fig:scenario_2}a). 
When focusing on Fig. \ref{fig:scenario_2}b), we see that the parameters of Hyd~2 contribute most to the first three eigenvectors. 
Especially in the first eigenvector, all Hyd~2 parameters show up, out of which parameter 8 and 12, being the $k_\text{hyd}$ and $k_\text{is}$ parameter, are most pronounced. 
A small contribution of parameter 13 ($k_\text{sec}$ of Hyd~2) can be observed in the first three eigenvectors. 
While only some parameters of Hyd~1 and 3 appear with small contributions in the first eigenvector, we can identify a dominating contribution of parameter 19 ($k_\text{is}$) in eigenvector 2 and 3. 
Those findings are in good agreement with the global sensitivity metrics in Fig. \ref{fig:scenario_2}c). 
This suggests that the parameters of Hyd~2 are most sensitive in scenario 2, followed by a pronounced sensitivity of the $k_\text{is}$ parameter of Hyd~3.    

\begin{figure}[htb]
	\centering
	\includegraphics[scale=0.4]{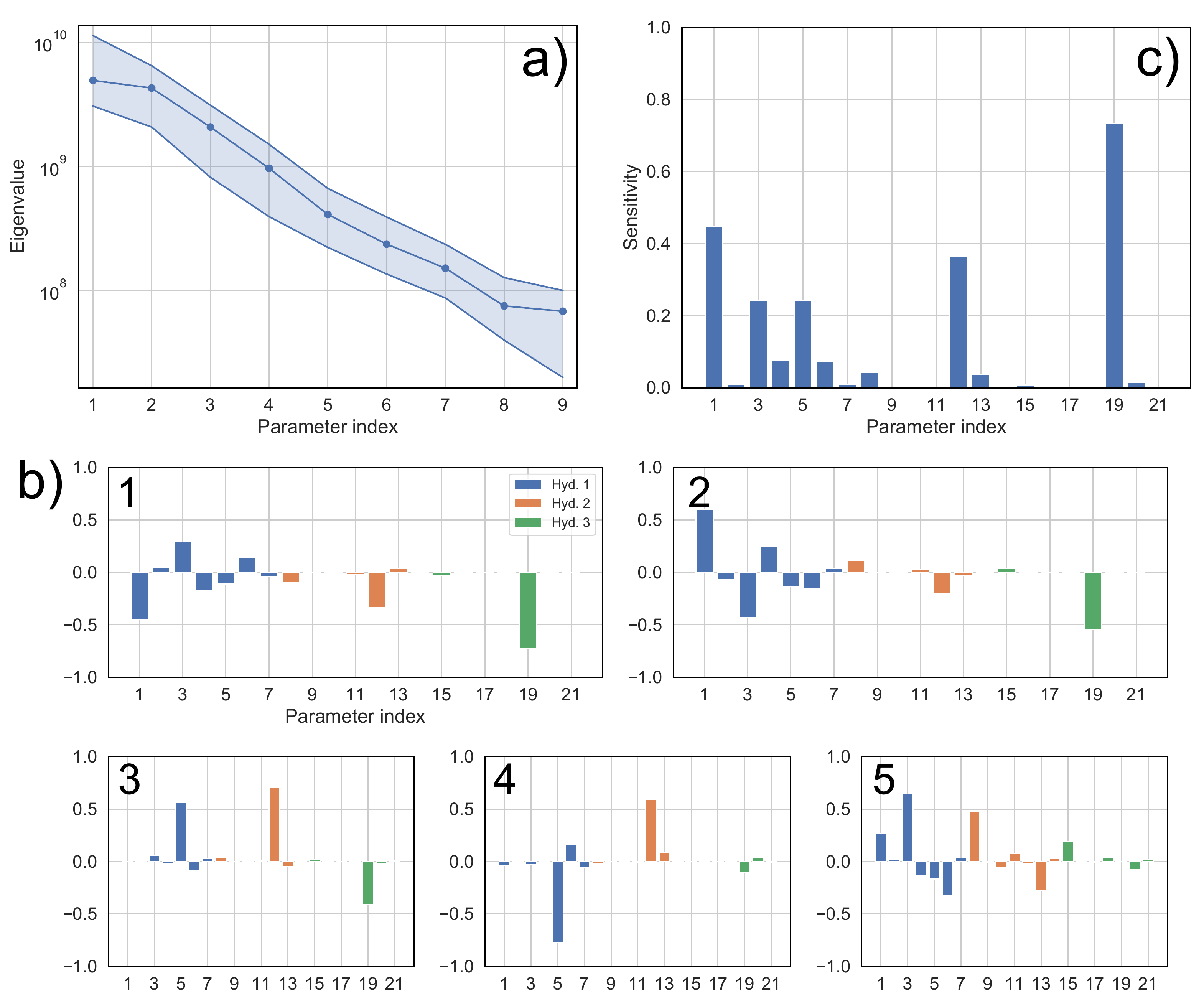}
	\caption{Results of the active subspace method applied to scenario 3 of the Kerschbaum spring LuKARS model. a) The eigenvalue decay as related to the first 9 eigenvectors. b) The parameter sensitivities of all 21 calibration parameters.  c) The first 5 eigenvectors of the 21-dimensional parameter space.}
	\label{fig:scenario_3}
\end{figure}

The results of scenario 3 are plotted in Fig. \ref{fig:scenario_3}. 
The eigenvalue decay, shown in Fig. \ref{fig:scenario_3}a), does not show a pronounced spectral gap, indicating that no active subspace can be found in the presented scenario. 
Therefore, similarly to scenario 1, we focus on all eigenvectors located within the range of one order of magnitude of the eigenvalue decay (here: 5).
In contrast with scenario 2, the contributing parameters in the dominant eigenvectors are related to Hyd~1 (Fig.~\ref{fig:scenario_3}b). 
However, parameters 12 and 19 also appear with a significant contribution in these eigenvectors. 
Parameter 19 has the largest contribution in eigenvector 1.
The most pronounced parameter contributions of Hyd~1 are attributed to parameter 1 and 3, which are the $k_\text{hyd}$ and $E_\text{max}$ parameters. 
A similar pattern can be observed in the global sensitivity metrics in Fig. \ref{fig:scenario_3}c), where the highest parameter sensitivity is assigned to parameter 19. 
Furthermore, we can observe that all parameters of Hyd~1 are sensitive, out of which the sensitivities of parameters 1, 3, and 5 ($k_\text{hyd}$, $E_\text{max}$ and $k_\text{is}$) are most pronounced. 
Parameter 12 is the most sensitive parameter of Hyd~2.

\subsection{Real case}
\begin{figure}[htb]
	\centering
	\includegraphics[scale=0.4]{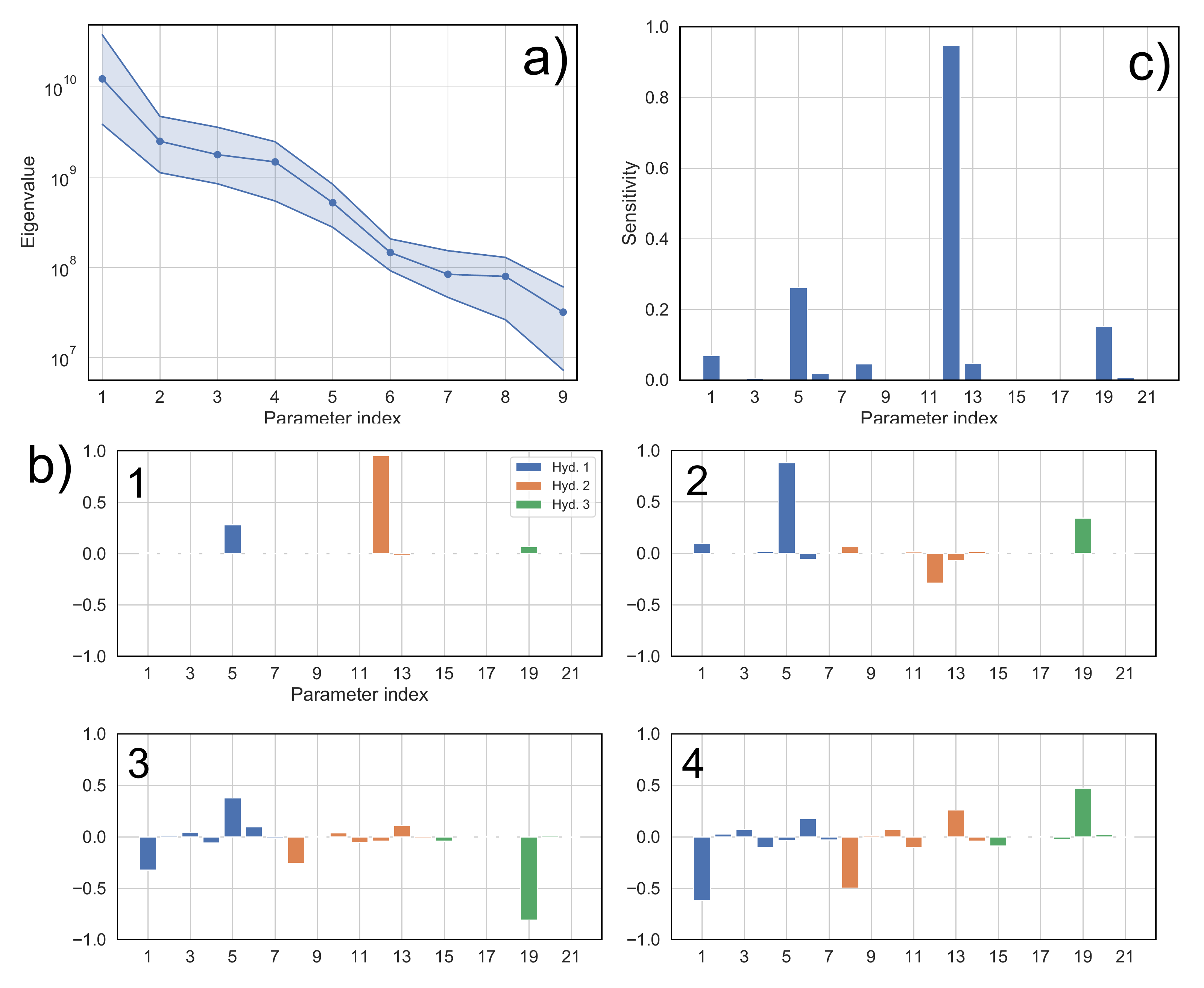}
	\caption{Results of the active subspace method applied to the real case scenario of the Kerschbaum spring LuKARS model. a) The eigenvalue decay as related to the first 9 eigenvectors. b) The parameter sensitivities of all 21 calibration parameters. c) The first 4 eigenvectors of the 21-dimensional parameter space.}
	\label{fig:scenario_rc}
\end{figure}

Finally, we show the results of the active subspace method applied to the real case scenario of the Kerschbaum spring LuKARS model in Fig. \ref{fig:scenario_rc}. 
Spectral gaps after the first and fourth eigenvalue (Fig. \ref{fig:scenario_rc}a) are considered indicative for the presence of 1-dimensional and a 4-dimensional active subspaces. 
In this regard, we recognize a large contribution of parameters 5, 12, and 19 to the first three eigenvectors (Fig. \ref{fig:scenario_rc}b). 
These represent the $k_\text{is}$ parameters of Hyd~1, Hyd~2, and Hyd~3. 
Also, we can identify a ranking between the related hydrotopes of these parameters in the first eigenvector. 
This ranking shows a decreasing order of the $k_\text{is}$ parameters from Hyd~2 to Hyd~1 to Hyd~3. 
In contrast, parameters 1, 8, and 15, having a remarkable contribution in eigenvectors 2 to 4, show a different ranking pattern. 
These parameters are related to $k_\text{hyd}$ of Hyd~1 - Hyd~3 and have a decreasing eigenvector contribution from Hyd~1 to Hyd~2 to Hyd~3. 
It should be noted that parameter 15 only has a small contribution in eigenvector 3 and 4. 
Another notable parameter group is related to parameters 6, 13, and 20, representing the $k_\text{sec}$ parameters of Hyd~1 - Hyd~3. Like the $k_\text{hyd}$ parameters, the $k_\text{sec}$ parameters show up in eigenvectors 2 to 4 with similar contributions in Hyd~1 and Hyd~2 (parameters 6 and 13) and with a minor contribution in Hyd 4 (eigenvector 4). 
The global sensitivity metrics obtained from the active subspaces are shown in Fig. \ref{fig:scenario_rc}c). 
It highlights that the $k_\text{is}$ parameters of all hydrotopes are the most sensitive parameters in the real case scenario. 
Further sensitive parameters are the $k_\text{hyd}$ parameters of Hyd~1 and Hyd~2 and the $k_\text{sec}$ parameters of Hyd~1 - Hyd~3, thus reflecting the parameter contributions found in the first 4 eigenvectors. \par  

\begin{figure}[htbp]   
	\centering
	\includegraphics[scale=0.5]{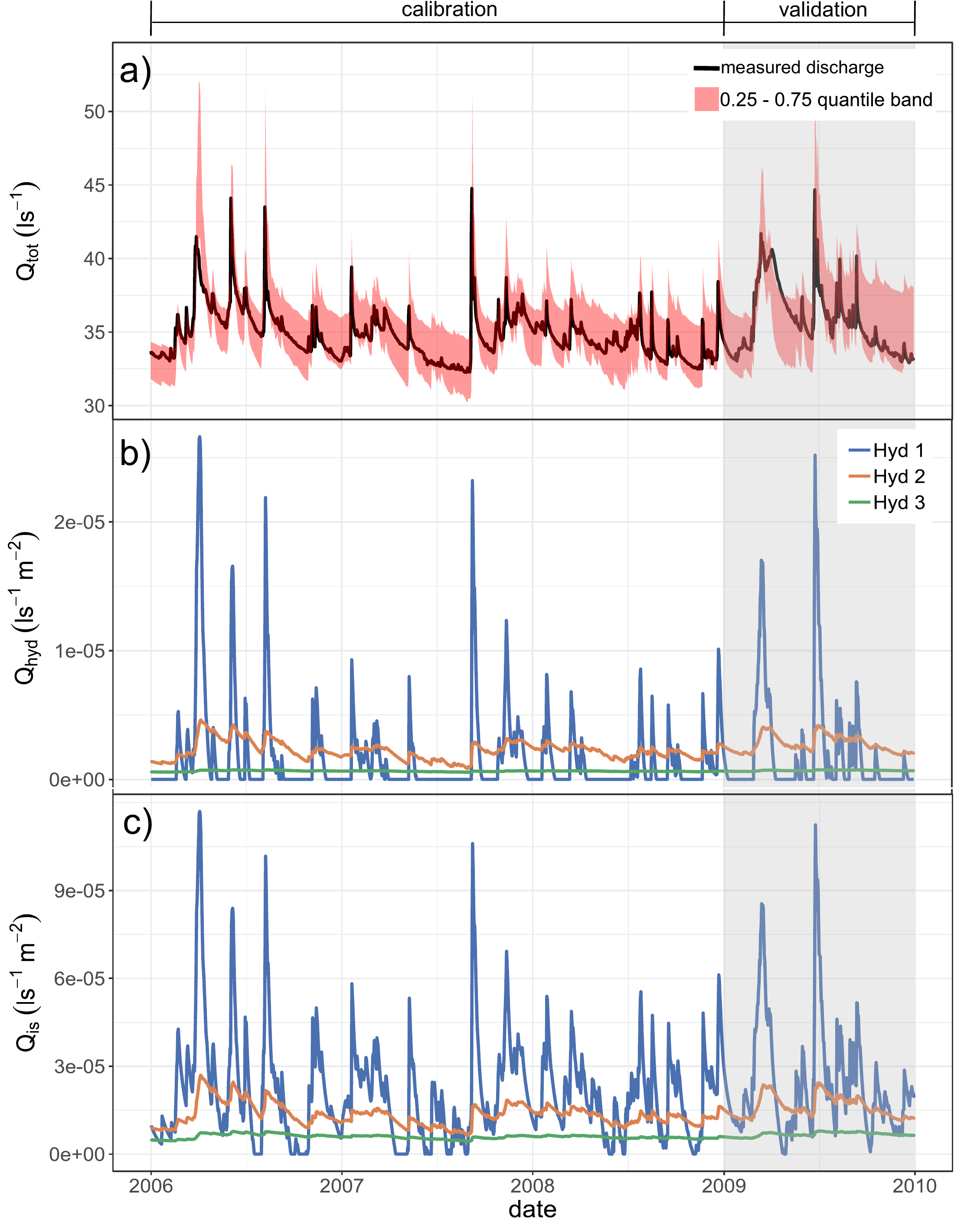}
	\caption{Modeling results of LuKARS using 1000 random samples of the posterior from the 4-dimensional active subspace model of the Kerschbaum recharge area derived by \citet{TeixeiraParente.2019}. a) The 0.25 - 0.75 quantile band of the 1000 forward runs and the observed discharge of the Kerschbaum spring for the calibration (01/2006 - 12/2008) and the validation period (01/2009 - 12/2009). b) and c) The specific quickflow contributions and the specific recharge from each single hydrotope, respectively, using the parameters of the mean of the push-forward run distribution.}
	\label{fig:model_results}
\end{figure}

\begin{table}[htb]
	
	\centering
	\caption{Results of the active subspace method applied to the three synthetic scenarios (1-3) and the real case (rc).}
	\begin{tabular}{p{0.5cm}p{6.0cm}p{3.0cm}p{3.5cm}}
		\hline\noalign{\smallskip}
		No. & spectral gap & sensitive parameters & parameters in relevant eigenvectors \\
		\hline\hline\noalign{\bigskip}
		1 & no spectral gap, eigenvectors considered over one order of magnitude of eigenvalue decay &  $k_\text{hyd}$, $k_\text{is}$ and $k_\text{sec}$ of all hydrotopes	&  $k_\text{hyd}$, $k_\text{is}$ and $k_\text{sec}$ of all hydrotopes     \\	\noalign{\bigskip}
		2 & spectral gap after third eigenvalue	& $k_\text{hyd}$, $k_\text{is}$ and $k_\text{sec}$ of Hyd~2 and $k_\text{is}$ of Hyd~3	& $k_\text{hyd}$, $k_\text{is}$ and $k_\text{sec}$ of Hyd~2 and $k_\text{is}$ of Hyd~3  \\	
		\noalign{\bigskip}
		3 & no spectral gap, eigenvectors considered over one order of magnitude of eigenvalue decay	& $k_\text{hyd}$, $E_\text{max}$ and $k_\text{is}$ of Hyd~1, $k_\text{is}$ of Hyd~2 and 3	& $k_\text{hyd}$, $E_\text{max}$ and $k_\text{is}$ of Hyd~1, $k_\text{is}$ of Hyd~2 and 3   \\
		\noalign{\bigskip}
		rc & spectral gap after fourth eigenvalue	& $k_\text{is}$ of all hydrotopes, $k_\text{hyd}$ of Hyd~1 & $k_\text{is}$ of all hydrotopes, $k_\text{hyd}$ of Hyd~1      \\
		\hline
	\end{tabular}
	\label{tab:scenario_results}
\end{table}
%Using the 4-dimensional subspace of the real case scenario, \citet{TeixeiraParente.2019} performed Bayesian inversion to find a posterior distribution of the model parameters. 
In Fig. \ref{fig:model_results}a), we show the measured discharge time series and the 0.25 to 0.75 quantile band as a result of 1000 forward model runs using random samples of the posterior distribution found by \citet{TeixeiraParente.2019}. 
The calibration period was from 01/2006 to 12/2008 and the validation period from 01/2009 to 12/2009. 
We observe that the uncertainty band obtained from the 4-dimensional subspace is well-centered around the observed time series.
Further, we plot the specific quickflow contributions ($Q_\text{hyd}$) and the specific interstorage discharge ($Q_\text{is}$, groundwater recharge) originating from each hydrotope in Fig. \ref{fig:model_results}b) and Fig. \ref{fig:model_results}c), respectively. 
We observe that Hyd~1 has the highest variability of specific discharge when looking at the quickflow and the recharge. 
Moderate discharge dynamics were found for Hyd~2. 
The specific discharge contributions of Hyd~3 are low and do not show a considerable variability. 

\begin{table}[htbp]
	\centering
	\label{tab:posterior}
	\caption{Posterior means and standard deviations of the physical parameters found by \citet{TeixeiraParente.2019} for the LuKARS model of the Kerschbaum spring recharge area.}
	\begin{tabular}{cccc}
		\hline\noalign{\smallskip}
		No. & phys. par.          & mean                           & std. \\
		\hline\hline\noalign{\smallskip}
		$1$ & $k_{\text{hyd},1}$  & $3.07 \times 10^2$             & $2.34 \times 10^2$ \\
		$2$ & $E_{\text{min},1}$  & $29.86$                        & $11.57$ \\
		$3$ & $E_{\text{max},1}$  & $44.49$                        & $12.90$ \\
		$4$ & $\alpha_1$          & $1.17$                         & $0.26$ \\
		$5$ & $k_{\text{is},1}$   & $5.18 \times 10^{-2}$ & $3.98 \times 10^{-3}$ \\
		$6$ & $k_{\text{sec},1}$  & $0.17$                         & $0.22$ \\
		$7$ & $E_{\text{sec},1}$  & $47.78$                        & $12.95$ \\
		\hline\noalign{\smallskip}
		$8$  & $k_{\text{hyd},2}$ & $70.62$                        & $55.81$ \\
		$9$  & $E_{\text{min},2}$ & $60.46$                        & $11.27$ \\
		$10$ & $E_{\text{max},2}$ & $1.20 \times 10^{2}$           & $16.14$ \\
		$11$ & $\alpha_2$         & $0.82$                         & $0.21$ \\
		$12$ & $k_{\text{is},2}$  & $4.52 \times 10^{-3}$ & $1.61 \times 10^{-4}$ \\
		$13$ & $k_{\text{sec},2}$ & $2.03 \times 10^{-2}$          & $3.23 \times 10^{-2}$ \\
		$14$ & $E_{\text{sec},2}$ & $1.76 \times 10^2$             & $25.99$ \\
		\hline\noalign{\smallskip}
		$15$ & $k_{\text{hyd},3}$ & $25.94$                        & $21.75$ \\
		$16$ & $E_{\text{min},3}$ & $95.71$                        & $14.18$ \\
		$17$ & $E_{\text{max},3}$ & $2.06 \times 10^2$             & $20.23$ \\
		$18$ & $\alpha_3$         & $0.43$                         & $0.14$ \\
		$19$ & $k_{\text{is},3}$  & $6.35 \times 10^{-4}$ & $1.69 \times 10^{-5}$ \\
		$20$ & $k_{\text{sec},3}$ & $6.21 \times 10^{-3}$          & $1.07 \times 10^{-2}$ \\
		$21$ & $E_{\text{sec},3}$ & $3.85 \times 10^2$             & $37.48$ \\
		\hline
	\end{tabular}
\end{table}

\begin{figure}     
	\centering
	\includegraphics[width=\textwidth]{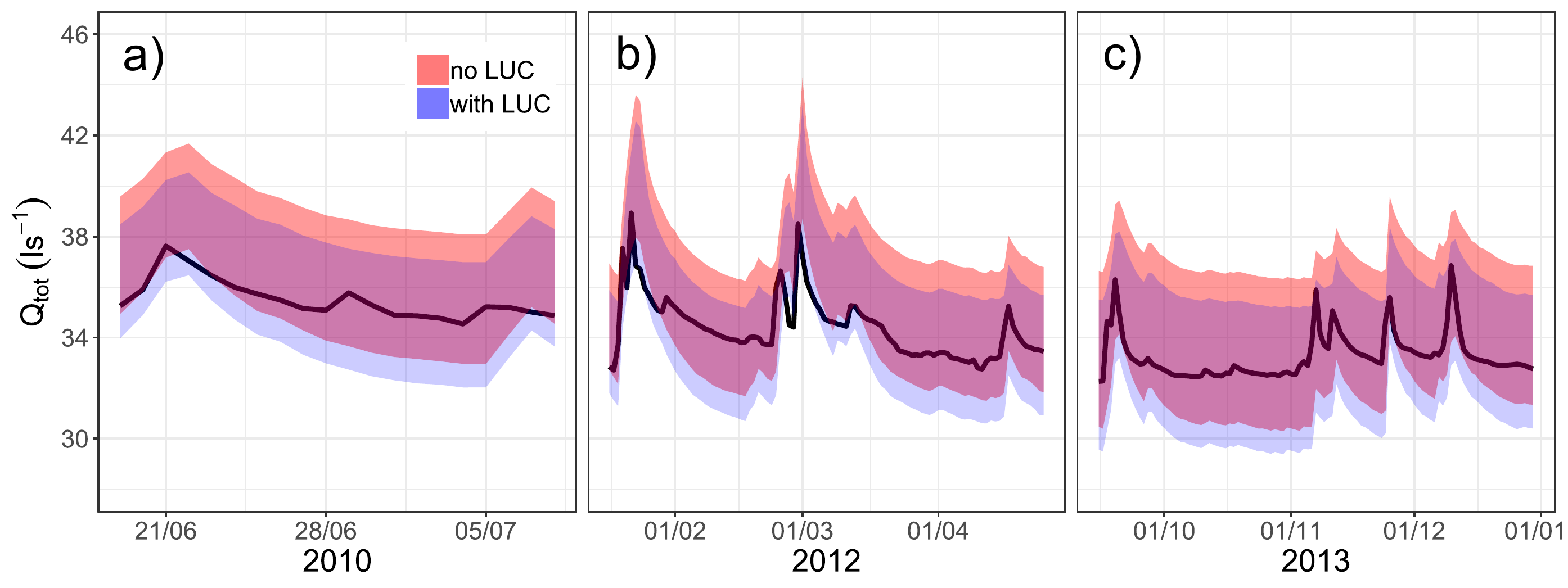}
	\caption{0.25 - 0.75 quantile bands obtained from 1000 forward runs of the 4-dimensional active subspace model when considering the increase of the dolomite quarries to 7\% of spatial share in the Kerschbaum recharge area (blue quantile band) and when not considering it (red quantile band). The quantile bands are shown for three distinct periods in a) 2010, b) 2012 and c) 2013.}
	\label{fig:luc_results}
\end{figure}

Furthermore, we want to investigate if the posterior found by \citet{TeixeiraParente.2019} for the 4-dimensional subspace during the calibration period (2006 to 2008, Table \ref{tab:posterior}) is able to reproduce the hydrological impacts of the increased area of the dolomite quarries that occured after the year 2009.
\cite{Bittner.2018} found that this land use change provoked a decrease of the mean spring discharge in the Kerschbaum spring. 
Therefore, we plot the uncertainty bands (0.25 - 0.75 quantile) for the forward simulations when considering and when not considering land use change in Fig. \ref{fig:luc_results}. 
Given that the measured time series included considerable data gaps from 2010 to 2013, it was not possible to validate the model results for the entire time series. 
However, when considering three distinct periods (20/06 - 09/07/2010, 18/01 - 26/04/2012, and 16/09 - 31/12/2013), we were able to validate the model results for high and low flow periods.
Regarding the model not considering this land use change (\ref{fig:luc_results}), we can observe that the measured time series is mostly closer to the 0.25 quantile than to the 0.75 quantile.
So, the uncertainty band of this model mainly overestimates the measured time series.
In contrast, we can observe that the model including the increased dolomite quarries is well centered around the measured discharge time series.   

%figures/

\section{Discussion}
\subsection{Synthetic scenarios}
In scenario 1, our aim was to investigate the simplest hydrotope setting having three hydrotopes with equal properties and the same area coverage.
The specific objective was to see if the active subspace method identifies this hydrotope equality by showing that all hydrotopes and their parameters are similarly informed by the data.
In fact, regarding the results shown in Fig. \ref{fig:scenario_1}b), we can verify that the parameters of each hydrotope are similarly informed with only negligible variations. 
In particular in the first eigenvector, the contributions of all 7 hydrotope parameters are similar between all hydrotopes. 
Moreover, the gaps observed in the eigenvalue decay (Fig. \ref{fig:scenario_1}a) are not significant, suggesting that no low-dimensional subspace exists in the proposed scenario. 
This is due to the fact that all hydrotopes have the same properties in terms of their physical characteristics as well as their area coverage. 
Given this similarity, there are no significant directions in which the parameters of one hydrotope are more informed than others.

Our interest when generating scenario 2 was to confirm the conclusions drawn from scenario 1 and to investigate what happens if we have three hydrotopes with the same phyiscal properties but with different area coverages. 
The parameters appearing in the relevant eigenvectors (eigenvector 1 - 3, Fig. \ref{fig:scenario_2}b) are related to the hydrotope covering the largest area in this synthetic scenario, Hyd~2. 
Hence, the area covered by each hydrotope directly affects how the measured discharge data (here the synthetic discharge time series of scenario 2, see Fig. \ref{fig:ts_scenarios}) informs the parameter space of LuKARS. 
The most informed paramters are $k_\text{is}$ and $k_\text{hyd}$ of Hyd~2 as well as $k_\text{is}$ of Hyd~3. 
This fact is further confirmed by the global sensitivity metrics shown in Fig. \ref{fig:scenario_2}c), where the mentioned parameters appear as the most sensitive ones. 
The reason why the $k_*$ parameters are most sensitive is that they constrain the amount as well as the variability of the three types of hydrotope discharges ($Q_\text{hyd}$, $Q_\text{is}$ and $Q_\text{sec}$). 
Therefore, to reduce the data misfit function, it is important that the discharge components of the largest hydrotopes approximate the evaluation data as good as possible. 
Scenario 2 highlights that the active subspace method is able to identify the impact of the area covered by a hydrotope on the spring discharge.

We generated scenario 3 to investigate with the active subspace method how the physical properties of each hydrotope affect the parameter sensitivities. 
Although Hyd~1 is not the largest hydrotope in terms of area coverage, its parameters show the largest contributions in the relevant eigenvectors (Fig. \ref{fig:scenario_3}b). 
This can be related to the dynamic discharge behavior, i.e. discharge variability, it reflects due to its physical properties. 
Hyd~1, as in the real case, has shallow soils and a coarse-grained soil texture, thus showing quick and variable discharge responses to precipitation and melt events. 
%Similar to scenario 2, this can be attributed to the data misfit function, which strives to match the entire variability of the observed discharge time series.
Moreover, it is interesting to observe that parameter 19 is more sensitive than parameter 1. Parameter 19 is $k_\text{is}$ of Hyd~3 and constrains the water transfered from the hydrotopes to the linear baseflow storage. 
%The baseflow was determined by \citet{Bittner.2018} and exhibits a low-variable and relatively constant discharge dynamic. 
We argue that $k_\text{is}$ of Hyd~3 has the highest sensitivity because Hyd~3 is the largest hydrotope in this scenario (covering 40\% of the entire recharge area). 
Considering that the baseflow storage ($E_\text{b}$) is controlled by two discharge coefficients, $k_\text{b}$ and $k_\text{is}$, as well as that $k_\text{b}$ was not included in the active subspace framework, $k_\text{is}$ of Hyd~3 considerably controls the discharge variability of the baseflow. 
Parameter 1 ($k_\text{hyd}$ of Hyd~1), in contrast, constrains the quickflow from the hydrotope with the most variable discharge behavior and hence accounts for simulating the discharge variability of the (synthetic) spring discharge. 
In conclusion, our findings highlight that the active subspace method is able to identify the relation between discharge data and i) the model structure, ii) the area of a hydrotope and iii) the physical properties of a catchment. 

\subsection{Real case}
The parameters appearing in the first eigenvector of the real case example (Fig. \ref{fig:scenario_rc}b) are the $k_\text{is}$ parameters of each hydrotope.
Similar to the results of scenario 3, we argue that $k_\text{is}$ of Hyd~2 has the highest contribution in the first eigenvector because Hyd~2 is the largest hydrotope in the catchment (covering 56\% of the entire recharge area) and its $k_\text{is}$ parameter controls the dynamics of the baseflow. 
This area dependence, however, cannot be observed when looking at $k_\text{is}$ of Hyd~1 and 3. 
Although Hyd~3 covers more space in the recharge area, the eigenvector contribution of $k_\text{is}$ from Hyd~1 is higher than this of Hyd~3. 
The reasons for that can be found in the hydrophysical properties of these hydrotopes and the resulting more dynamic discharge behavior of Hyd~1 (i.e. discharge variability) as compared to Hyd~3. 
%Hyd~3 has a much more homogeneous hydrological response to precipitation events than Hyd~1 and Hyd~2, and 
From a hydrological point of view, Hyd~3 has the highest storage capacity and the most fine-grained soil texture, leading to a better water retention capacity. 
This causes that the discharge processes originating from Hyd~3, i.e. $Q_\text{hyd}$ and $Q_\text{is}$, are attenuated and more homogeneous as compared to Hyd~1 and Hyd~2 (see Fig. \ref{fig:model_results}b for $Q_\text{hyd}$ and Fig. \ref{fig:model_results}c for $Q_\text{is}$ of each hydrotope). 
Hence, the discharge data does not inform the discharge parameters of Hyd~3 as much as those of Hyd~1 and Hyd~2. 
It concludes that Hyd~1 plays a more significant role in matching the observed discharge dynamics as compared to Hyd~3.
The same effect is also highlighted in eigenvectors 3 and 4, where $k_\text{hyd}$ of Hyd~1 is more pronounced as compared to $k_\text{hyd}$ of Hyd~3. \par   
%These observations can further be related to the effect of the data misfit function, which tries to match the variability of the observed spring discharge.
%This effect can be attributed to the data misfit function, where we try to fit the entire dynamics of a measured discharge signal.
%Using a different objective evaluation function here might lead to a different result.\par 
Generally, the way the hydrotope parameters are informed by spring discharge data with respect to the physical properties of and the area covered by a hydrotope is indicative about the robustness of the hydrotope approach. 
It highlights that, although the model is lumped and not physically based, the hydrotopes’ parameters can be inferred by discharge observations and a physical explanation can be found by their sensitivities.
In addition, it shows that the active subspace method can be used to investigate model structure uncertainties since its results are affected by the characteristics of an investigated catchment (here: the hydrotope properties).   \par 
As shown by \citet{TeixeiraParente.2019}, the prior intervals of those parameters contributing to the dominant eigenvectors are most constrained by the discharge data. 
In our case, the results tell us that the prior distribution of the $k_\text{is}$ parameters (see Table \ref{param_intervals}) becomes most constrained by the measured discharge in the Kerschbaum spring (see standard deviations in Table \ref{tab:posterior}). 
Hence, the parameter uncertainties of those parameters contributing to the dominant eigenvectors can be most reduced. 
Moreover, the points discussed can further be validated by the global sensitivity analysis shown in Fig. \ref{fig:scenario_rc}c): we can see that $k_\text{is}$ of Hyd~2 is the most sensitive parameter of the model. 
As discussed for the first eigenvector, this fact is plausible, since $k_\text{is}$ of Hyd~2 constrains the discharge contribution to the baseflow storage and the baseflow itself. 
The results of the sensitivity analysis further confirm that the parameters responsible for simulating the higher discharge dynamics of Hyd~1 ($k_\text{is}$ and $k_\text{hyd}$) are more sensitive than the $k_*$ parameters of Hyd~3, although Hyd~3 covers larger areas in the catchment as Hyd~1. 
We can interpret that linear combinations of the parameters present in the relevant eigenvectors (they show the dominant directions over which an active subspace can be spanned) are highly sensitive. 
Given this information about the sensitive parameter directions (eigenvector information) as well as that we can compute a global sensitivity metrics from the approximated eigenpairs (see section \ref{S:active_subsp}), we argue that the active subspace helps finding a physical explanation of the parameter sensitivities. \par 

Using 1000 random samples of the posterior from the 4-dimensional subspace derived by \citet{TeixeiraParente.2019}, we can observe that that the uncertainty band (0.25-0.75 quantile band, Fig. \ref{fig:model_results}) adequately matches the observed time series.
Furthermore, the hydrotope-specific discharge contributions, as displayed by $Q_\text{hyd}$ and $Q_\text{is}$ in Fig. \ref{fig:model_results}b) and c), reflect the expected discharge variability with respect to the physical properties of each individual hydrotope.
Hence, the active subspace method recognizes catchment-specific characteristics when identifying the most informed parameter directions. 
When it comes to the simulation of land use changes, we can identify that the 4-dimensional subspace model is able to capture the hydrological impacts of the increase of the area used for dolomite mining.
We draw this conclusion since the uncertainty band obtained from the model considering that land use change is well-centered around the measured time series (see Fig. \ref{fig:luc_results}).
These facts emphasize that using samples of the identified active subspaces leads to acceptable model results yet reducing the 21-dimensional to a 4-dimensional problem. 
Moreover, the model result uncertainties of the reduced parameter space model (represented by the quantile band) are small with respect to the mean discharge of the Kerschbaum spring ($<$ 10\%). 
All in all, our results highlight the applicability of the proposed methodology for parameter dimension reduction and uncertainty quantification.

%\subsection{Implications for other lumped karst aquifer models}
%The results discussed in the previous sections provide valuable indications for the application of the active subspace method to other lumped karst aquifer models. Typical lumped models for karst aquifers include buckets to represent the different compartments of a karst aquifer, i.e. the epikarst, matrix and conduits. These can include different combinations of buckets to simulate different connections between different compartments and varying numerical approaches. Using the active subspace method can thus be promising to investigate the realism of a given model structure and to justify possible modifications. Moreover, this method can help to evaluate the relative importance of any considered part of a karst system, which contributes to the reduction of uncertainties related to the model structure. In case of an existing active subspace, this method can be used in a parameter estimation framework in which the prior ranges of the most informed parameters become constrained and the parameter uncertainties can be quantified. By identifying a realistic model structure and deriving the different important compartments of a karst system, this framework can also support decision-making in water management contexts.  

\section{Conclusion}
\label{S:Conclusion}
In this work, we applied the active subspace method to the LuKARS rainfall-discharge model \cite{Bittner.2018} for four different scenarios (three synthetic and a real case), all with a 21-dimensional parameter space. 
Therefore, we provided a framework in which we adapt the physical model parameters of LuKARS to make the active subspace method applicable for the model while maintaining the physical constraints between each hydrotope.
Our aim was to investigate how much each parameter of the model is informed by the discharge data and how different model setups affect the results of the active subspace analysis. 
We found that both the relative area covered by a defined hydrotope and its discharge variability can have a major impact on parameter sensitivities and on how model parameters are informed by discharge data. 
In particular, the scenarios with no or minor variations of the catchment characteristics, i.e. scenario 1 and 2, show that model parameters are similarly informed by data. 
For these scenarios, we found that the area covered by a hydrotope has the largest impact on how model parameters are informed. 
In contrast, the scenarios with more marked differences of the catchment properties, i.e. scenario 3 and the real case, show pronounced differences in parameter sensitivities. 
We highlight that the hydrological behavior, i.e. the discharge variability of a hydrotope, can play a more significant role than the area coverage in informing a model parameter. 
This relationship is strongly related to how much the measured discharge signal depends on the discharge variability of a regarded hydrotope. 
In total, we found that the discharge coefficients ($k_*$ parameters) were the most sensitive parameters in all scenarios.  
%Generally, the more pronounced the differences of physical hydrotope properties are in a well-defined catchment, the more informed (sensitive) get the discharge-constraining parameters of those hydrotopes displaying a high discharge variability in comparison to the area it covers.
Using three different synthetic hydrotope scenarios of the Kerschbaum springshed, we highlighted the potential of using the active subspace method for investigations related to model structure and model parameter uncertainties. 
Given the different ways how different hydrotopes and their parameters are informed by the discharge data, the results of the active subspace method provide evidence that the hydrotope-based modeling approach is robust, although having a 21-dimensional parameter space in the real case scenario (Kerschbaum recharge area, Austria).
Based on the 0.25 - 0.75 quantile band obtained from 1000 randomly sampled forward runs, we showed that the calibrated model based on the identified 4-dimensional subspace is able to reasonably reproduce the measured discharge of the Kerschbaum spring.
Also, the 4-dimensional subspace model is able to simulate the impacts of the land use changes which occured in the Kerschbaum recharge area, highlighting the general applicability of LuKARS for land use change impact studies.

In conclusion, our results provide valuable indications for the application of the active subspace method to other lumped karst aquifer models.  
We suggest to apply the active subspace method to other lumped karst aquifer models with different representations of karst systems.
Considering different model structures, it is worth investigating which parameters can be effectively calibrated given a set of observations. 
This can help to reduce model output uncertainties and to potentially justify using a lumped parameter model with a high-dimensional parameter space.

\section*{Acknowledgment}
Daniel Bittner and Gabriele Chiogna refer to the Interreg Central Europe project PROLINE-CE funded by ERDF. 
Gabriele Chiogna further acknowledges the support of the Stiftungsfonds für Umweltökonomie und Nachhaltigkeit GmbH (SUN).
Additional financial support for Barbara Wohlmuth, Steven Mattis and Mario Teixeira Parente was provided by the German Research Foundation (DFG, Project WO 671/11-1).
All authors acknowledge support from the UNMIX project (12.05) funded by TUM IGSSE.
The water works in Waidhofen a.d. Ybbs kindly provided the orthophotos as well as the relevant discharge, precipitation and temperature data. 
We also thank Roland K\"ock (BOKU, Wien) for providing preliminary information about the hydrotopes in the catchment area.

\bibliography{references}

%\end{linenumbers}

\end{document}